# High oxygen pressure floating zone growth and crystal structure of the metallic nickelates $R_4Ni_3O_{10}$ (R=La, Pr)

Junjie Zhang,[1,2,3]* Hong Zheng,[1] Yu-Sheng Chen,[4] Yang Ren,[5] Masao Yonemura,[6,7] Ashfia Huq,[8] and J. F. Mitchell[1]*

[1]Materials Science Division, Argonne National Laboratory, Argonne, IL 60439, United States

[2]Materials Science and Technology Division, Oak Ridge National Laboratory, Oak Ridge, TN 37831, United States

[3]Institute of Crystal Materials, Shandong University, Jinan, Shandong 250100, China

[4]ChemMatCARS, The University of Chicago, Argonne, IL 60439, United States

[5]X-ray Science Division, Advanced Photon Source, Argonne National Laboratory, Argonne, IL 60439, United States

[6]Institute of Materials Structure Science, High Energy Accelerator Research Organization, 1-1 Oho, Tsukuba, Ibaraki 305-0801, Japan

[7]Sokendai (The Graduate University for Advanced Studies), Shirakata 203-1, Tokai, Naka 319-1106, Japan

[8]Neutron Scattering Science Directorate, Oak Ridge National Laboratory, Oak Ridge, Tennessee 37831, United States

*email: junjie@sdu.edu.cn; mitchell@anl.gov

**ABSTRACT**:

Single crystals of the metallic Ruddlesden-Popper trilayer nickelates $R_4Ni_3O_{10}$ (R=La, Pr) were successfully grown using an optical-image floating zone furnace under oxygen pressure ($pO_2$) of 20 bar for $La_4Ni_3O_{10}$ and 140 bar for $Pr_4Ni_3O_{10}$. A combination of synchrotron and laboratory x-ray single crystal diffraction, high-resolution synchrotron x-ray powder diffraction and measurements of physical properties revealed that $R_4Ni_3O_{10}$ (R=La, Pr) crystallizes in the monoclinic $P2_1/a$ (Z=2) space group at room temperature, and that a metastable orthorhombic phase (*Bmab*) can be trapped by post-growth rapid cooling. Both $La_4Ni_3O_{10}$ and $Pr_4Ni_3O_{10}$ crystals undergo a metal-to-metal transition (MMT) below room temperature. In the case of $Pr_4Ni_3O_{10}$, the MMT is found at ~157.6 K. For $La_4Ni_3O_{10}$, the MMT depends on the lattice symmetry: 147.5 K for *Bmab* vs. 138.6 K for $P2_1/a$. Lattice anomalies were found at the MMT that, when considered together with the pronounced dependence of the transition temperature on subtle structural differences between *Bmab* and $P2_1/a$ phases, demonstrates a not insignificant coupling between electronic and lattice degrees of freedom in these trilayer nickelates.

# 1. INTRODUCTION

Quasi-2D transition metal oxides with strong electron correlations have attracted considerable attention due to their collective properties such as high-temperature superconductivity in cuprates [1-5], colossal magnetoresistance in manganites [6], and metal-insulator transitions in nickelates [7]. Ordering of charge, spin and orbital are important phenomena underlying these exotic physical properties [1-5,8]. In particular, charge order, or charge density wave (CDW) order, has been established as an instability of the underdoped cuprate superconductors [9]. Charge order has also been found in non-copper transition metal oxides containing mixed valence states, such as $Mn^{3+}/Mn^{4+}$ [10], $Ni^{2+}/Ni^{3+}$ [11], $Ni^{1+}/Ni^{2+}$ [12], $Fe^{2+}/Fe^{3+}$ [13], and $Co^{2+}/Co^{3+}$ [14]. One notable characteristic of charge order in these latter systems is their insulating ground state, instead of the metallic state found in cuprates [8]. The metallic, mixed-valent Ruddlesden-Popper trilayer nickelates, $R_4Ni_3O_{10}$ (R=La, Pr and Nd), nominally contain $Ni^{2+}/Ni^{3+}$; the average valence state is 2.67+, which can be thought as 0.67 holes doped into a background of $Ni^{2+}$. Such hole doping, if mapped onto the cuprates, would lie far into the overdoped regime, which is generally regarded as conventional Fermi liquid possessing no collective electronic order. However, Peng *et al.* recently discovered charge order in the overdoped cuprate $(Bi,Pb)_{2.12}Sr_{1.88}CuO_{6+\delta}$ without nesting or pseudogap features [15]. It is thus relevant to establish if the ground state of these metallic trilayer nickelates expresses CDW order to better delimit what is a growing body of work exploring the connections between nickelates to cuprates, most dramatically expressed through the recent discovery of superconductivity in thin films of $Nd_{0.8}Sr_{0.2}NiO_2$ [16].

The trilayer nickelates $R_4Ni_3O_{10}$ (R=La, and Pr) have been reported to undergo metal-to-metal transitions (MMTs, ≈140 K for $La_4Ni_3O_{10}$ and ≈160 K for $Pr_4Ni_3O_{10}$ and $Nd_4Ni_3O_{10}$) [17-24]. The transition was ascribed by Zhang and Greenblatt to CDW-driven instabilities [18]. Later, Seo *et al.* published one-electron, tight-binding band structure (at the extended Hückel level), predicting a CDW nesting vector of $q$=0.3$a^*$ or 0.3$b^*$ [25,26]. However, direct evidence of CDW formation, *e.g.,* the observation of superlattice lines by x-ray, neutron or electron diffraction, has not been reported in polycrystalline samples [18,21,27]. An obvious next step would be study of single crystals. However, a challenge to crystal growth originates from the fact that $La_4Ni_3O_{10}$ decomposes to $La_3Ni_2O_7$ and NiO before it melts, as can be seen from the La-Ni-O phase diagram in Ref. [28]. The advent of high $p$O$_2$ floating-zone crystal growth capabilities has overcome this problem, and crystals of $R_4Ni_3O_{10}$ have been grown by us [12,29,30] and others [23]. Indeed, using these crystals in a combined x-ray and neutron scattering study, we have recently discovered compelling evidence of the simultaneous formation of intertwined charge- and spin density waves in the metallic oxide, $La_4Ni_3O_{10}$ [31].

Despite considerable focus on these materials, the crystal structure of $R_4Ni_3O_{10}$ (R=La, Pr, and Nd) above the MMT still remains under debate. Four different space groups have been proposed in the literature: (*i*) *Bmab* (alternate setting of *Cmce*), as reported by Tkalich *et al* [17], Ling *et al* [21], and V.I. Voronin *et al* [32]. (*ii*) *Imm*2: Zhang *et al.* reported *Fmmm* from powder x-ray diffraction, but their electron diffraction patterns show that the space group symmetry is lower than *F* or *C*, and the highest symmetry space group possible is *Imm*2 [18]. (*iii*) $P2_1/a$ (Z=4): Olafsen *et al.* reported $P2_1/a$ (a=5.3675(2) Å, b=5.4548(2) Å, c=27.433(1) Å, $\beta$=90.312(2)°) for $Nd_4Ni_3O_{9.85}$ based on refinement of powder neutron diffraction data at 298 K [33]. Later, Nagell *et al.* [34], Kumar *et al.*,[35] Huangfu *et al.*[36], and Li *et al.*[24] refined their powder diffraction data using this space group. Notably, the above three structures were obtained from Rietveld refinements on powder diffraction data, which suffer from significant peak broadening arising from anisotropic strain and stacking faults [34,37]. (*iv*) $P2_1/a$ (Z=2): Huangfu *et al.* reported $P2_1/a$ (a=5.3771(5) Å, b=5.4549(3) Å, c=14.028(2) Å, $\beta$ =100.910(9)°) from lab x-ray single crystal diffraction of $Pr_4Ni_3O_{10}$ at 293 K [23]. Whether $La_4Ni_3O_{10}$ and $Nd_4Ni_3O_{10}$ crystallize in the same $P2_1/a$ (Z=2) is an open question.

In this contribution, we report single crystal growth of $R_4Ni_3O_{10}$ (R=La, Pr) and establish their crystal structures above the MMT. The outline of the present paper is as follows. First, we present the successful growth of bulk single crystals of $R_4Ni_3O_{10}$ (R=La and Pr) using the high-pressure optical-image floating zone technique. The obtained stable phases with increasing $p$O$_2$ follow along a progression with $n$ in $La_{n+1}Ni_nO_{3n+1}$. Second, we investigate the physical properties including resistivity, magnetic susceptibility and heat capacity of $R_4Ni_3O_{10}$, where we find $La_4Ni_3O_{10}$ exhibits an unexpected phase transition at $T_{MMT}$=147.5 K in addition to the 138.6 K transition reported in literature. We attribute these transitions to metastable orthorhombic and stable monoclinic phases, respectively. In contrast, $Pr_4Ni_3O_{10}$ shows a single MMT at 157.6 K, consistent with a single-phase sample. Third, we determine the crystallographic structure above the MMT using synchrotron and lab x-ray single-crystal diffraction. Both $La_4Ni_3O_{10}$ with $T_{MMT}$=138.6 K and $Pr_4Ni_3O_{10}$ crystallize in the monoclinic space group $P2_1/a$ (Z=2) reported by Huangfu for $Pr_4Ni_3O_{10}$ [23], while the $La_4Ni_3O_{10}$ with $T_{MMT}$=147.5 K belongs to the orthorhombic structure *Bmab*. Fourth, we discuss annealing experiments on bi-phasic samples to address the question of the true thermodynamic phase of $R_4Ni_3O_{10}$ at room temperature, *Bmab* or $P2_1/a$. Finally, we summarize our findings.

## 2. EXPERIMENTAL

**2.1 Solid-State Reaction**. Precursors for crystal growth were synthesized via standard solid-state reaction techniques. Stoichiometric amounts of $La_2O_3$ (Alfa Aesar, 99.99%) or $Pr_6O_{11}$ (Alfa Aesar, 99.99%) and NiO (Alfa Aesar, 99.99%) were thoroughly ground. For $Pr_4Ni_3O_{10}$, an excess of 0.5-1.5% NiO was added to compensate the loss during growth due to volatilization during growth at high $pO_2$ [38]. The mixture was then loaded into a Pt crucible and heated in air from room temperature to 1050 °C at a rate of 3 °C/min, allowed to dwell for 24 h, then furnace-cooled to room temperature. The solid was then reground and sintered twice at 1050 °C using the procedures mentioned above. Powder x-ray diffraction showed a phase mixture consisting of $La_4Ni_3O_{10}$, $La_3Ni_2O_7$, $La_2NiO_4$, $La_2O_3$ and NiO (see Figure S1).

**2.2 High-$pO_2$ Crystal Growth**. Crystals of $R_4Ni_3O_{10}$ (R=La, Pr) were successfully grown using a vertical optical-image floating-zone furnace designed for operation at elevated gas pressure (150 bar Model HKZ, SciDre GmbH, Dresden). Precursor powders of $La_4Ni_3O_{10}$ were hydrostatically pressed into polycrystalline rods (length=100 mm, diameter=8 mm) and sintered for 24 h at 1400 °C to make a dense rod. $La_4Ni_3O_{10}$ crystals were grown directly from the sintered rod at $pO_2$ = 20 bar using a 3 kW Xenon arc lamp to heat the zone. A similar procedure was applied to $Pr_4Ni_3O_{10}$ but severe cracking in the feed rod during growth resulted in melt zone loss. By reducing the sintering temperature to 1100 °C and making a less dense rod, the cracking issue was avoided. $Pr_4Ni_3O_{10}$ crystals were grown at $pO_2$ = 140 bar through two steps using a 5 kW Xenon arc lamp. The first step was a fast pass (30-50 mm/h) to improve density. This was followed by a second growth at the same pressure with slow travelling rate, 5 mm/h. During growth, a flow rate of 0.1 L/min of oxygen was maintained. Feed and seed rods were counter-rotated at 27 rpm and 20 rpm, respectively, to improve zone homogeneity.

**2.3 Single-Crystal Structural Determination**. Single crystal x-ray diffraction data were collected with a Bruker APEX2 area detector using lab x-ray at room temperature ($\lambda$=0.71073 Å) and using synchrotron radiation at 200 K at Beamline 15-ID-D at the Advanced Photon Source, Argonne National Laboratory. Single crystals were mounted to the tip of glass fibers and collected using a Bruker D8 diffractometer. Indexing was performed using Bruker APEX2 software [39]. Data integration and cell refinement were performed using SAINT, and multi-scan absorption corrections were applied using the SADABS program [39]. The structure was solved by direct methods and refined with full matrix least-squares methods on $F^2$. All atoms of monoclinic $R_4Ni_3O_{10}$ (R=La, Pr) and La atoms of orthorhombic $La_4Ni_3O_{10}$ were modeled using anisotropic ADPs, and the refinements converged for $I > 2\sigma$ ($I$), where I is intensity of reflections and $\sigma$ ($I$) is standard deviation. Calculations were performed using the SHELXTL crystallographic software package [40]. Details of crystal parameters, data collection and structure refinement are summarized in Table I. Selected bond distances (Å) and angles (°) are given in Table II. Further details of the crystal structure investigations may be obtained from the joint CCDC/FIZ Karlsruhe online deposition service: https://www.ccdc.cam.ac.uk/structures/ by quoting the deposition number CSD 1999906-1999908.

**2.4 Sample Annealing.** Under flowing oxygen, as-grown single crystals of $La_4Ni_3O_{10}$ were heated to 1000 °C at a rate of 200 °C/h, held for 12 hours, and then cooled to 200 °C at a rate of 30 °C/h, and finally to room temperature by turning off the furnace. High $pO_2$ annealing was performed in a specially constructed annealing furnace (Model AHSO, SciDre GmbH, Dresden) under 20 bar oxygen pressure. Samples were held for 2 hours at 950 ℃ or 1050 °C, and then quenched to room temperature.

**2.5 Powder X-ray Diffraction (PXRD)**. Room temperature PXRD was performed on a PANalytical X'Pert PRO diffractometer using Cu K$\alpha$ radiation ($\lambda$=1.5418 Å). High resolution PXRD data were collected at Beamline 11-BM in the range $0.5° \leq 2\theta \leq 28°$ with a step size of 0.001° and counting time of 0.1 s per step and a wavelength of $\lambda$=0.459003 Å for $Pr_4Ni_3O_{10}$ and $\lambda$=0.414579 Å for $La_4Ni_3O_{10}$. Samples were prepared by loading pulverized crystals into a $\Phi$0.5 mm quartz capillary. The quartz capillary was then loaded inside a $\Phi$0.8 mm Kapton capillary for installation on a magnetic sample base used by the beamline sample changer. The sample was spun continuously at 5600 rpm during data collection. An Oxford Cryostream 700 Plus $N_2$ gas blower was used to control temperature below room temperature. Diffraction patterns were recorded at room temperature first (1 h scan) and then on cooling from 296 to 100 K (10 min per scan, cooling rate of 0.3 K per min). The same sample of $La_4Ni_3O_{10}$ before and after annealing in flowing oxygen were measured at room temperature in the range $0.5° ≤ 2θ ≤ 50°$ with a step size of $0.001°$ and counting time of 0.1 s per step and a wavelength of $λ= 0.414167$ Å. Temperature dependent diffraction data above room temperature were collected at 50-200 (at 50 °C intervals), 220-380 (at 10 °C intervals), 430-550 (at 10 °C intervals), 600 and 650 °C with temperature controlled by a hot air blower. Data were analyzed with the Rietveld method using GSAS-II software [41]. The background at each temperature was fit using Chebyshev polynomial (14 terms). Other refined parameters include scale factor, sample displacement perpendicular to beam, phase ratio, lattice parameters, atomic positions, isotropic atomic displacement parameters (all like atoms were grouped together), and profile shape parameters. Isotropic domain size and generalized microstrain models were used. Parametric sequential fitting was performed for temperature dependent data.

**2.6 Neutron Powder Diffraction (NPD)**. Temperature dependent NPD data were collected on a time-of-flight (TOF) Super High Resolution Powder Diffractometer, Super HRPD [42,43] at the Material and Life science Facility (MLF) in the Japan Proton Accelerator Research Complex (J-PARC). The powder sample was installed in a cylindrical vanadium cell of dimensions 6 mm in diameter, 55 mm in height, and 100 μm in thickness. All diffraction data were collected using all detectors in the range $10º < 2θ < 172º$. The diffraction data from the backscattering bank ($150° < 2θ < 172º$) were used for the analysis. Structural refinements were performed using *Z*-Rietveld software [44,45].

**2.7 Oxygen Content.** Oxygen content of $R_4Ni_3O_{10}$ (R=La, Pr) was determined by reduction in a 4% $H_2/N_2$ mixture on a thermogravimetric analysis (Mettler Toledo Model TGA/DSC 1). Pulverized samples (~100 mg) taken from the crystal boule were placed into a 150 μL $Al_2O_3$ crucible and heated at a rate of 10 °C/min from room temperature to 900 °C, held for 5 hours, and then cooled at 10 °C/min to room temperature (see Figure S2). Multiple blanks had been run previously to establish stability and buoyancy correction.

**2.8 Heat Capacity**. Heat capacity measurements were performed on a Quantum Design PPMS in the temperature range of 1.8-300 K. Apiezon-N vacuum grease was employed to fix crystals to the sapphire sample platform. The specific heat contribution from sample holder platform and grease was determined before mounting sample and subtracted from the total heat capacity.

**2.9 Magnetic Susceptibility**. Magnetic susceptibility measurements were performed on single crystals using a Quantum Design MPMS3 SQUID magnetometer. Single crystals were attached to a quartz holder using a minute amount of adhesive. ZFC-W (Zero-field cooling with data collected on warming), FC-C (field cooling and data collected on cooling) and FC-W (field cooling and data collected on warming) data with magnetic field H∥ab plane and H⊥ab were collected between 1.8 and 300 K under an external field of 0.4 T. In the ZFC-W protocol the sample was cooled in zero field to 10 K at a rate of 35 K/min and then to 1.8 K at a rate of 2 K/min, and DC magnetization recorded on warming (2 K/min). In the FC-C and FC-W protocols, the magnetization was recorded (2 K/min) in a fixed field of 0.4 T. Isothermal field-dependent magnetization at 1.8 K and 300 K with magnetic field H∥ab plane and H⊥ab were measured in a field range of ±7 T following field cooling (10 K/min) under 7 T.

**2.10 Electrical resistivity**. Resistivity of $La_4Ni_3O_{10}$ single crystals was measured using a four-terminal method while $Pr_4Ni_3O_{10}$ single crystals was measured using a six-terminal method [46,47] with contacts made by depositing gold pads. Temperature was controlled using the Quantum Design PPMS in the temperature range of 1.8−300 K.

## 3. RESULTS AND DISCUSSION

**High $pO_2$ Crystal Growth**. We first explored phase formation as a function of oxygen pressure by melting and rapid cooling materials with a starting composition of La:Ni=4:3 (see Figure S1). At pO₂= 0.21 bar O₂, $La_4Ni_3O_{10}$ decomposes to $La_3Ni_2O_7$, NiO and O₂ at ~1400 K based on the phase diagram reported by Zinkevich and Aldinger [22,28]. The lack of a tie line to a liquid phase precludes growth of a single crystal from the melt at ambient pressure. Considering the high valence state of Ni (nominal +2.67), a highly oxidizing environment is expected to help stabilize the target phase. Figure 1 presents the lab powder x-ray diffraction patterns of the phases obtained by rapid cooling the melt at various $pO_2$ with $La_2NiO_4$, $La_3Ni_2O_7$, $La_4Ni_3O_{10}$ and $LaNiO_3$ as references. The data confirm that high $pO_2$ is crucial for the formation of various Ruddlesden-Popper nickelates and that the stable phases with increasing pressure follow along the progression with $n$ in $La_{n+1}Ni_nO_{3n+1}$, *i.e.*, $La_2NiO_4$ was obtained at low $pO_2$,

La$_3$Ni$_2$O$_7$ was the majority phase at ~14 bar, La$_4$Ni$_3$O$_{10}$ was obtained when $p$O$_2$ is in the range of 16-30 bar (see Figure S3), and LaNiO$_3$ was the major phase at and above 50 bar. We have already reported the successful growth of metallic LaNiO$_{3-\delta}$ at 40 and 130 bar O$_2$ [48,49], as has Guo *et al* [50] and Dey *et al.* [51] Figure 1b schematically illustrates the oxygen pressure stability range for different phases. As expected, this is in line with nominal Ni valence states.

To optimize the $p$O$_2$ for R$_4$Ni$_3$O$_{10}$ (R=La and Pr) crystal growth, it is critical to use as high $p$O$_2$ as possible to suppress the formation of single-layer R$_2$NiO$_4$ and bilayer R$_3$Ni$_2$O$_7$, but low enough to avoid the perovskite phase, RNiO$_3$. We found a trace of LaNiO$_3$ at $p$O$_2$ = 25 bar for La$_4$Ni$_3$O$_{10}$ growth in lab x-ray powder diffraction patterns, thus we adopted 20 bar $p$O$_2$ for crystal growth of La$_4$Ni$_3$O$_{10}$. The same procedure was applied to the case of Pr-Ni-O, and we found Pr$_2$NiO$_4$ formed at low pressure and Pr$_4$Ni$_3$O$_{10}$ above 100 bar O$_2$. We were unable to stabilize either the bilayer Pr$_3$Ni$_2$O$_7$ or the perovskite PrNiO$_3$ below the 150 bar limit of our furnace. The missing Pr$_3$Ni$_2$O$_7$ may reflect a very narrow pressure region of stability or an inversion of the stability ranges of Pr$_3$Ni$_2$O$_7$ and Pr$_4$Ni$_3$O$_{10}$ vis-à-vis the La counterparts. PrNiO$_3$ lies at $p$O$_2$ higher than 150 bar, as we recently reported for the single crystal growth of this compound at ~300 bar $p$O$_2$ [52]. For the crystal growth of Pr$_4$Ni$_3$O$_{10}$, we employed oxygen pressure of 140 bar.

To obtain high-quality single crystals of R$_4$Ni$_3$O$_{10}$ (R=La and Pr), we also investigated other growth parameters including traveling rates, feeding rates, rotation speeds, and post-growth cooling rates. We found that fast travelling rates, e.g., 30-50 mm/h, introduce second phases (*e.g.*, La$_2$NiO$_4$) during growth, consistent with reports of disordered intergrowths of different *n* members of Ruddlesden-Popper phases [18]. Traveling rates of 4-6 mm/h for the seed were found to be acceptable. The rotation speeds of feed and seed rods are related to the growth interface between solid and liquid, and a planar or slightly convex interface shape with respect to the growth crystal is reported to be desirable [53]. As demonstrated in Ref. [54] for floating zone growth of Y$_2$Ti$_2$O$_7$, the shape of the growth interface changes from convex to less convex, and finally to concave with the speed of rotation of the crystals decreasing from 30 to 7 rpm. In our crystal growth, we applied rotation speeds of 27 and 20 rpm for feed rod and seed, respectively, to achieve a stable melt zone and a convex growth front. Post-growth cooling rate is an important parameter in obtaining thermodynamically stable phase and/or trapping metastable phases if a material undergoes first-order structural phase transition(s) on cooling. As reported by Nagell *et al*.[37], La$_4$Ni$_3$O$_{10}$ undergoes a series of phase transitions on cooling, from tetragonal to orthorhombic to monoclinic. We utilized two cooling rates after crystal growth: (i) Keep traveling with fixed power, i.e., slow cooling, or (ii) Power off immediately after separating feeding and seeding parts, i.e., rapid cooling. (i) was used to obtain the thermodynamically stable phase, and (ii) was used to trap a metastable phase.

A typical boule of La$_4$Ni$_3$O$_{10}$ is shown in Figure 2a. The growth direction is parallel to the *ab* plane. Crystals with shiny facets were cleaved from the as-grown boule, as shown in Figure 2b. Figure 2c shows a cleaved La$_4$Ni$_3$O$_{10}$ crystal, ~ 4.5 mm in length. Diffraction patterns at various positions along the length of this crystal measured in transmission using synchrotron x-rays (λ=0.1173 Å) at beamline 11-ID-C, Advanced Photon Source, are shown in Figures 2d-m. The observation of the same diffraction pattern from position *d* to *l* indicates a large single crystal (i.e. coherent growth) with good crystallinity. The oxygen content of specimens randomly extracted from as-grown La$_4$Ni$_3$O$_{10}$ and Pr$_4$Ni$_3$O$_{10}$ crystals was measured to be 9.98(1) and 10.05(1), respectively (see Figure S2). With deviation of ≤ 5 parts per thousand, both as-grown La$_4$Ni$_3$O$_{10}$ and Pr$_4$Ni$_3$O$_{10}$ are nominally stoichiometric.

**Physical properties.** Resistivity, magnetic susceptibility and heat capacity were measured on as-grown R$_4$Ni$_3$O$_{10}$ (R=La, Pr) crystals. La$_4$Ni$_3$O$_{10}$ was found to exhibit two different phase transition temperatures depending on the post-growth cooling: (i) $T_{\mathrm{MMT}}$=138.6 K using slow cooling rate after growth, and (ii) $T_{\mathrm{MMT}}$=147.5 K by rapid cooling the boule after growth. We will show below that the 138.6 K transition comes from a monoclinic ($P2_1/a$) phase and the 147.5 K transition from an orthorhombic (*Bmab*) phase (see **Crystal Structure Determination**). In contrast, Pr$_4$Ni$_3$O$_{10}$ shows a single $T_{\mathrm{MMT}}$ at 157.6 K, which is in the $P2_1/a$ phase. We will refer to these phases by their space group symmetry throughout this discussion of their physical properties.

Figures 3(a-c) show the resistivity of R$_4$Ni$_3$O$_{10}$ as a function of temperature on warming. The in-plane resistivity drops with a decreasing of temperature from room temperature, indicating a metallic behavior. An anomaly was observed at 138.6 K, 147.5 K, and 157.6 K for $P2_1/a$-, *Bmab*-La$_4$Ni$_3$O$_{10}$ and Pr$_4$Ni$_3$O$_{10}$, respectively. To evaluate

the anisotropy, the resistivity of $Pr_4Ni_3O_{10}$ was measured using six-terminal method [46,47]. Its in-plane and out-of-plane resistivity at 300 K were measured to be 1.55 mΩ·cm and 5.96 mΩ·cm, respectively. The anisotropy at 300 K is 3.8, increasing to 11.7 at 2 K. Such an anisotropy is relatively small compared with other layered oxides such as manganites [46] and cuprates [55]. The temperature dependent behavior of $R_4Ni_3O_{10}$ around MMT resembles that of CDW materials such as chromium [56], purple bronzes [57], and rare-earth tritellurides [58].

The in-plane and out-of-plane magnetic properties of $R_4Ni_3O_{10}$ (R=La and Pr) were measured on single crystals under a magnetic field of 0.4 T (see Figures 3(d-f)) on warming and cooling. Because lanthanum is diamagnetic, we first look at the case of $P2_1/a$- and $Bmab$-$La_4Ni_3O_{10}$. Both the in-plane and out-of-plane magnetic susceptibility decrease on cooling above the MMT. Around the MMT, the $ab$-plane susceptibility sharply decreases with a decreasing temperature while the $c$-axis susceptibility reaches a minimum (at ~141 K for $P2_1/a$ and at ~149 K for $Bmab$), and then increases upon further cooling. Such behavior is neither Curie−Weiss nor Pauli [34], the latter expected of a metal. The lack of hysteresis between warming and cooling process indicates either a weakly first-order or a second-order transition. The magnetic susceptibility of $La_4Ni_3O_{10}$ is quite similar to CDW materials such as $K_{0.3}MoO_3$ [57], or spin density wave materials including $BaFe_2As_2$ [59]. We notice that $Bmab$-$La_4Ni_3O_{10}$ shows an irreversibility below 50 K that can be connected to a weak ferromagnetic component, corroborated by isothermal $M(H)$ data (see Figure S4). Such a ferromagnetic behavior has been reported previously in oxidized $La_4Ni_3O_{10}$ [20]. The magnetic susceptibility of $Pr_4Ni_3O_{10}$ is Curie-Weiss like due to the large contribution of the local paramagnetic moment from $Pr^{3+}$, which masks the underlying behavior of Ni moments. However, an anomaly is clearly seen in $d\chi/dT$ at 157.6 K, which is consistent with the anomaly seen in resistivity and heat capacity. Huangfu $et$ $al$. fitted $\chi$ to Curie-Weiss law below and above the transition, and reported a steplike feature in the magnetic susceptibility [23]. We performed high-resolution powder neutron diffraction on $R_4Ni_3O_{10}$ (R=La, Pr) to investigate whether the ground state is antiferromagnetic; however, no superlattice peaks were observed (See Figure S5). Later, our single crystal neutron diffraction measurements unambiguously revealed weak magnetic superlattice reflections in addition to main Bragg peaks, forming spin density waves below MMT [31].

The heat capacity of $P2_1/a$-, $Bmab$-$La_4Ni_3O_{10}$ and $P2_1/a$ $Pr_4Ni_3O_{10}$ is shown in Figures 3(g-i). The phase transitions are clearly indicated by a prominent anomaly on each curve. To estimate the change of entropy across MMT, we have phenomenologically fit the behavior above and below with a polynomial and subtracted this background. The difference between $C_p/T$ and the fit background is shown in Figure S6. Integrating the area under the resultant peak yields $\Delta S \sim 2.0$ J mole$^{-1}$ K$^{-1}$ for $P2_1/a$-$La_4Ni_3O_{10}$, $\Delta S \sim 2.1$ J mole$^{-1}$ K$^{-1}$ for $Bmab$-$La_4Ni_3O_{10}$, and $\Delta S \sim 2.7$ J mole$^{-1}$ K$^{-1}$ for $Pr_4Ni_3O_{10}$ (see Figure S6). In contrast, the entropy change is reported to be 1.5 J mole$^{-1}$ K$^{-1}$ for $P2_1/a$-$La_4Ni_3O_{10}$ by Kumar $et$ $al$ [35], and 2.0 J mole$^{-1}$ K$^{-1}$ for $Pr_4Ni_3O_{10}$ by Huangfu $et$ $al$.[23]. Inset of Figures 3(g,h) show the $C_p/T$ as a function of $T^2$ in the low temperature region. The heat capacity at low temperature is fit to the standard model of the specific heat in a nonmagnetic solid, $C_p/T=\gamma+\beta T^2$, where $\gamma$ is the electronic contribution and the $T^2$-term is from the lattice. The fit leads to $\gamma$=13.3 mJ mole$^{-1}$ K$^{-2}$ and $\beta$=3.7×10$^{-4}$ J mole$^{-1}$ K$^{-4}$ for $P2_1/a$-$La_4Ni_3O_{10}$ [31] and $\gamma$ =14.5 mJ mole$^{-1}$ K$^{-2}$ and $\beta$=3.4×10$^{-4}$ J mole$^{-1}$ K$^{-4}$ for $Bmab$-$La_4Ni_3O_{10}$. The large $\gamma$ values are consistent with their metallic ground states. The extracted $\beta$ values correspond to Debye temperatures of 450 [31] and 460 K, respectively. The obtained $\gamma$ values are in good agreement with the previous report by Kumar $et$ $al$.[35] and Wu $et$ $al$. [60]. For $Pr_4Ni_3O_{10}$, a Schottky anomaly is observed at low temperature, indicating a manifold of relatively low lying excited states probably due to crystal field splitting [61].

**Crystal Structure Determination**. As discussed in the introduction, there is considerable debate about the ground state crystal structure of $R_4Ni_3O_{10}$ materials. To address this lack of consensus, we turned to single crystal x-ray diffraction on our specimens. We were able to isolate from the as-grown boule three different kinds of $La_4Ni_3O_{10}$ crystals: single phase $Bmab$, single phase $P2_1/a$, and mixed phase $Bmab+P2_1/a$. For $Pr_4Ni_3O_{10}$, only $P2_1/a$ crystals were found. Using these crystals, we were able to establish that for both R=La and Pr the room temperature structure of $R_4Ni_3O_{10}$ is the monoclinic $P2_1/a$ structure with $Z$=2 reported by Huangfu $et$ $al$. [23]. Comprehensive details of the models tested and justification for selecting this monoclinic structure are provided in the Supplemental Information (see Synchrotron x-ray single crystal diffraction and Table S1). Here we summarize the results.

Figure 4 shows the structures of $Bmab$- and $P2_1/a$-$La_4Ni_3O_{10}$ in the $ab$, $bc$ and $ac$ plane. Although the atomic arrangement in both the orthorhombic and monoclinic structures is the same – trilayer perovskite-like blocks separated by a rocksalt layer—the unit cell of $Bmab$ consists of two such trilayers, while the $P2_1/a$ cell contains only one. For both $Bmab$- and $P2_1/a$-$La_4Ni_3O_{10}$, La atoms are in 9 and 12 coordination with oxygen atoms, and Ni atoms are surrounded by six oxygen atoms, although the bond lengths are different (see Table II). It is worth noting that the outer Ni of the trilayer has a larger distortion of its oxygen octahedron than the inner Ni. This is not unusual for multi-layer Ruddlesden-Popper phases, for instance the bilayer manganite $La_{1.2}Sr_{1.8}MnO_7$ [62]. Bond valence sum calculations [63] were performed to evaluate the valence states of Ni. The obtained values of inner and outer Ni differ (3.0 vs 2.7, see Table II), reflecting charge differentiation between them. Such an observation has been discussed in $Pr_4Ni_3O_{10}$ [23] and $Nd_4Ni_3O_{10}$ [24].

Figure 5 shows the heat capacity of a biphasic, as-grown sample of $La_4Ni_3O_{10}$, characterized by two anomalies that correspond to transitions in the $P2_1/a$ and $Bmab$ phases. Figure 6 shows the high-resolution synchrotron x-ray powder diffraction pattern and Rietveld refinement of biphasic $La_4Ni_3O_{10}$ at room temperature. The refinement converged to $R_{wp}$=12.1% and GOF=1.7 with a mass fraction of 42.0%wt for $Bmab$ and 58.0%wt for $P2_1/a$. The obtained lattice parameters are $a$=5.4193 Å, $b$=5.4722 Å, $c$=27.9656 Å for $Bmab$, and $a$= 5.419271 Å, $b$= 5.4717 Å, $c$= 14.2306 Å and $\beta$=100.712° for $P2_1/a$. The inset of Figure 6 shows the pattern and fit in the $Q$ range 2.255-2.277 Å$^{-1}$, where three peaks are clearly seen, two outer peaks from $P2_1/a$ and an inner one attributed to $Bmab$. Notably, the difference between $P2_1/a$ and $Bmab$ is so subtle that lab x-ray diffraction cannot resolve it (see Figure S3).

To link the crystal structures to their physical properties, we measured in situ high-resolution synchrotron x-ray powder diffraction on a pulverized biphasic crystal of $La_4Ni_3O_{10}$ on cooling from 300 K to 100 K. The evolution of the (200), (020) and (002) peaks as a function of temperature is shown in Figure S7. The shift of the peaks towards higher $2\theta$ on cooling indicates contraction of the unit cell above the MMT. No peak splitting or extra peaks were observed in the powder data below the MMT. Figure 7 presents the temperature dependence of the unit cell parameters $a$, $b$, $c$, $\beta$, and $V$ extracted from Rietveld refinements, as well as the calculated linear thermal expansion coefficients. Anomalies are seen in the lattice parameters and linear thermal expansion coefficients at 138.6 K and 147.5 K, corresponding to the monoclinic and orthorhombic phases, and reflecting lattice involvement in the MMTs. This response is particularly evident in the $b$ axis. For $Pr_4Ni_3O_{10}$, an anomaly at 157.6 K is observed (see Figures S8 and S9). It is understandable how such a subtle lattice response, ≈0.03%, was missed previously [17,21].

We performed in situ high-resolution synchrotron x-ray powder diffraction on a mixed-phase sample on warming from 321 to 923 K in air to explore for a thermally driven transition in $La_4Ni_3O_{10}$ from $P2_1/a$ to $Bmab$. A monoclinic to orthorhombic transition was not observed. Indeed, over this temperature interval, the monoclinic $\beta$ angle increases (Figure 7c), showing no tendency toward higher symmetry. However, the Rietveld refinements show that the $a$ and $b$ axes of the orthorhombic phase are converging with warming and upon extrapolation would become metrically equivalent at ~1160 K. Such an observation suggests the possibility of a transition to tetragonal phase, albeit at a significantly higher temperature than that reported by Nagell $et$ $al$. (973 K) [37] and Amow $et$ $al$. (758 K) [64]. If this is the case, it may indicate some difference in stoichiometry (presumably O content) between our sample and that of Nagell $et$ $al$. [37] The thermal expansion is anisotropic with smallest coefficient along the $b$ axis. Anomalies at ~550 K and ~750 K are observed, the former of which is consistent with the electronic transition seen in resistivity, magnetic susceptibility and differential thermal analysis reported by Kobayashi $et$ $al$. [19], and the latter of which is in agreement with the transition reported by Amow $et$ $al$ [64] but with a different and as-yet unknown origin.

**Thermodynamically stable phase of $R_4Ni_3O_{10}$.** An open question is the true thermodynamic phase of $R_4Ni_3O_{10}$ at room temperature, $Bmab$ or $P2_1/a$. Our biphasic crystals allow us to answer this question definitively. We annealed biphasic single crystals of $La_4Ni_3O_{10}$ under flowing oxygen at 1000 °C [65]. The heat capacity of the same specimen before and after annealing is presented in Figure 5. The peak centered at 147.5 K (associated with the $Bmab$ phase) becomes weaker after annealing, while the peak at 138.6 K (associated with the $P2_1/a$ phase) becomes more pronounced, reflecting a phase conversion from $Bmab$ to $P2_1/a$. Such an observation has been reproduced in multiple samples. Figures 8(a,b) show the diffraction patterns of as-grown biphasic $La_4Ni_3O_{10}$ and the same sample after annealing. Figure 8c presents the region around $Q$~2.265 Å$^{-1}$ before and after annealing in flowing oxygen with peaks normalized. The increase of the intensity attributable to the $P2_1/a$ phase after annealing demonstrates a thermally-driven conversion from $Bmab$ to $P2_1/a$ and confirms that the latter is the thermodynamic stable state. This is shown quantitatively in Table III: the mass fraction of $P2_1/a$ increases while that of $Bmab$ decreases, reflecting a conversion from $Bmab$ to $P2_1/a$. That $Bmab$ $Pr_4Ni_3O_{10}$ could be obtained by rapid cooling corroborates the metastability of the orthorhombic structure at room temperature (see Figure S10). We note that we cannot unequivocally rule out the possibility of a compositional change (e.g., O stoichiometry) during the annealing process. However, as shown in Table III, the lattice parameters are essentially unchanged (<1 part in 5000) compared to those before annealing, arguing that any such compositional change is most probably negligible.

## 4. CONCLUSION

In summary, we successfully obtained single crystals of $La_4Ni_3O_{10}$ with two crystal structures ($Bmab$ and $P2_1/a$) and $Pr_4Ni_3O_{10}$ ($P2_1/a$) using a high-pressure floating zone furnace. Crystal structures of $R_4Ni_3O_{10}$ (R=La, Pr) have been determined by combining synchrotron and lab x-ray single crystal diffraction, high-resolution synchrotron x-ray and physical properties including resistivity, magnetic susceptibility and heat capacity. Key outcomes of this work are: (1) The monoclinic $P2_1/a$ (Z=2) structure is established as the room temperature thermodynamic state of bot La- and $Pr_4Ni_3O_{10}$. For $La_4Ni_3O_{10}$, the $P2_1/a$ structure and a competing orthorhombic $Bmab$ phase are apparently nearly degenerate, shedding light on past controversies over the crystallography of this material; (2) A weak, anisotropic lattice response was observed at the MMT, substantiating a lattice coupling to the electronic degrees of freedom. (3) Despite the weak lattice response at the MMT itself, the link between lattice and the electronic bandstructure is not insubstantial, with the MMT differing in $La_4Ni_3O_{10}$ by ≈10 K (7%) between the $P2_1/a$ and $Bmab$ phases. Beyond this study, the availability of these $R_4Ni_3O_{10}$ (R=La, Pr) single crystals provides means to definitively test the CDW theory of the MMT [31], and to explore any putative analog between overdoped cuprates and nickelates in the framework of collective behavior exemplified by charge order.

## AUTHOR INFORMATION
Corresponding Author
junjie@sdu.edu.cn or mitchell@anl.gov

## ACKNOWLEDGMENTS

This work was supported by the US Department of Energy, Office of Science, Basic Energy Sciences, Materials Science and Engineering Division. This research has been supported in part by ORNL Postdoctoral Development Fund by UT-Battelle, LLC under Contract No. DE-AC05-00OR22725 with the U.S. Department of Energy. The work at Shandong University was supported by the Qilu Young Scholars Program of Shandong University, and the Taishan Scholars Program of Shandong Province. NSF's ChemMatCARS Sector 15 is supported by the Divisions of Chemistry (CHE) and Materials Research (DMR), National Science Foundation, under grant number NSF/CHE-1834750. Use of the Advanced Photon Source, an Office of Science User Facility operated for the U.S. Department of Energy (DOE) Office of Science by Argonne National Laboratory, was supported by the U.S. DOE under Contract No. DE-AC02-06CH11357. J. Z. thanks Dr. Saul H. Lapidus for his help with the high-resolution powder diffraction experiment at 11-BM, and Drs. Raphael P. Hermann, D. Phelan, A. S. Botana, M. R. Norman, J. W. Freeland, V. Pardo, W. E. Pickett, and S. Rosenkranz for fruitful discussions.

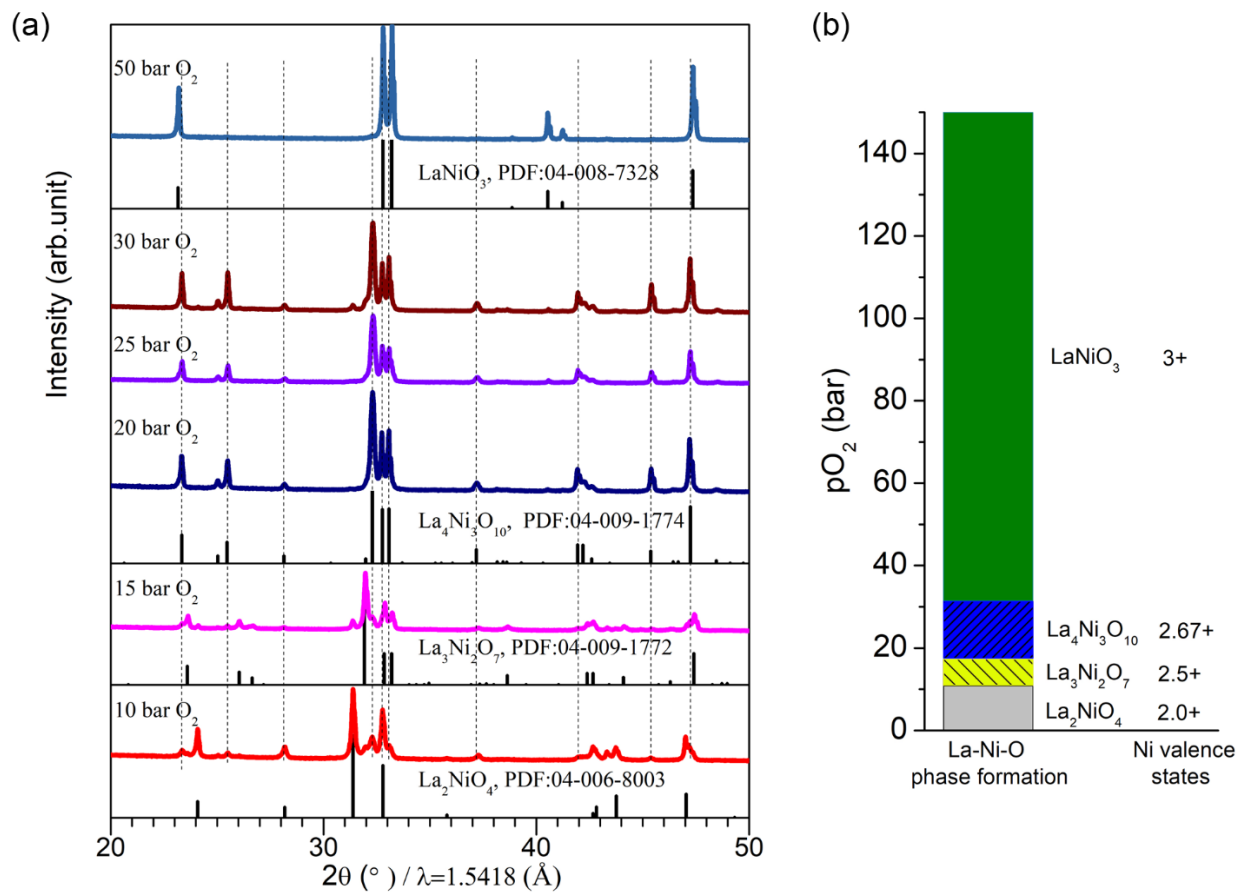

**Fig. 1** Rapid cooling products from a melt of $La_2O_3$: NiO=2: 3 at various oxygen pressure. (a) Powder x-ray diffraction patterns vs $pO_2$ with standard patterns of $La_{n+1}Ni_nO_{3n+1}$ ($n$=1, 2, 3, and $\infty$) from database shown as tick marks below the data; (b) Schematic drawing of empirical phase predominance as a function of $pO_2$.

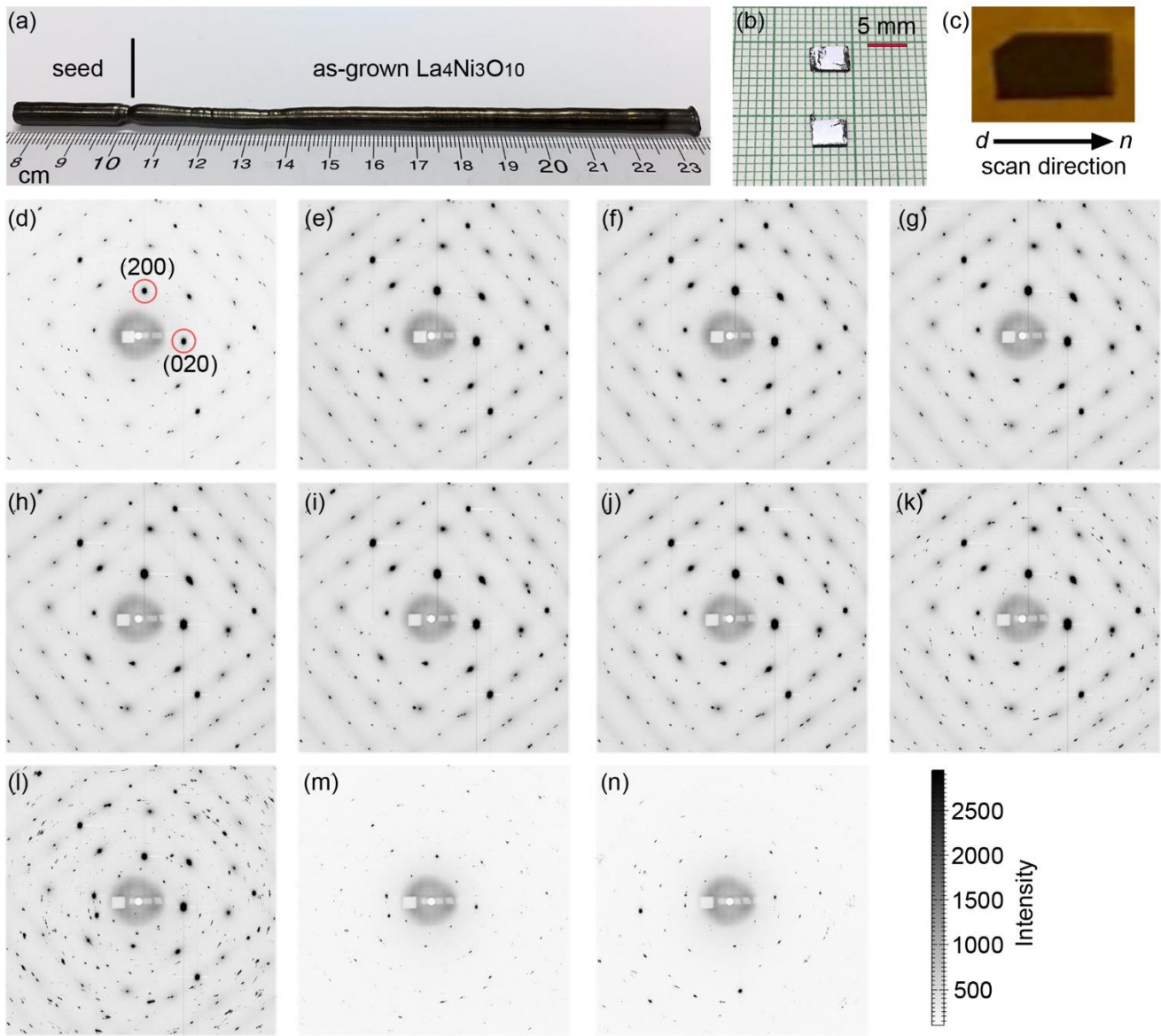

**Fig. 2** High-energy synchrotron x-ray single crystal diffraction from an as-grown $La_4Ni_3O_{10}$. (a) Photograph of as-grown boule with crystal growth direction parallel to the *ab* plane, (b) Cleaved $La_4Ni_3O_{10}$ crystals from (a), (c) a $La_4Ni_3O_{10}$ crystal ~ 4.5 mm in length attached on Kapton tape and its diffraction patterns at various positions (d-n) ($\lambda$=0.117 Å, 11-ID-C, APS; beam size 0.8x0.8 mm², scan step size 0.5 mm). Note that diffuse scattering is observed in the *hk*0 plane, reflecting some short-range deviation from the average structure. Note that the vertical lines in Figs. *e-l* are artifact caused by overexposure. The observation of multiple spots close together is due to twinning.

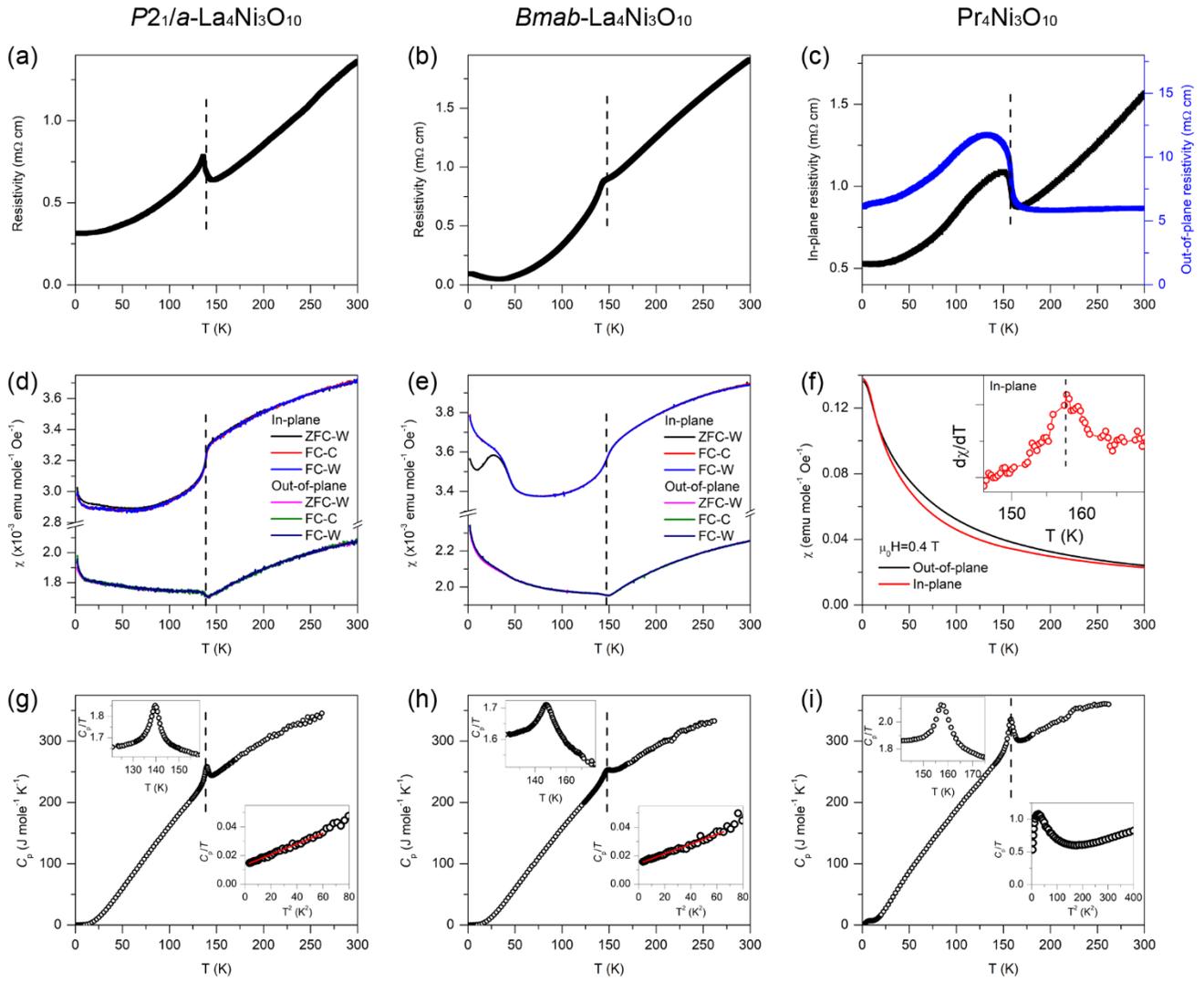

**Fig. 3** Physical properties of $P2_1/a$-, $Bmab$-La$_4$Ni$_3$O$_{10}$ and Pr$_4$Ni$_3$O$_{10}$. (a,d,g) Resistivity, magnetic susceptibility and heat capacity of $P2_1/a$-La$_4$Ni$_3$O$_{10}$. (b,e,h) Resistivity, magnetic susceptibility and heat capacity of $Bmab$-La$_4$Ni$_3$O$_{10}$. (c,f,i) Resistivity, magnetic susceptibility and heat capacity of Pr$_4$Ni$_3$O$_{10}$.

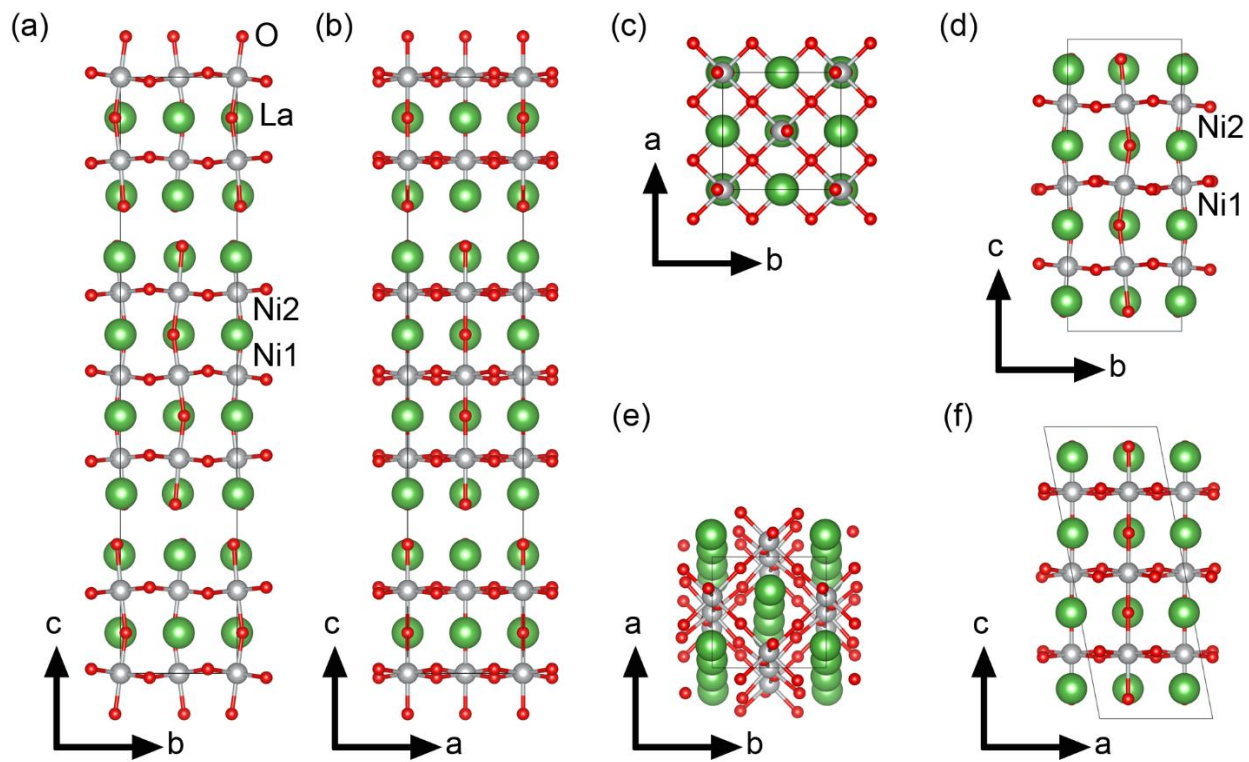

**Fig. 4** Crystal structures and coordination environments of Ni atoms in $R_4Ni_3O_{10}$ (R=La and Pr) from single crystal diffraction. (a-c) $Bmab$-$La_4Ni_3O_{10}$, (d-f) $P2_1/a$-$La_4Ni_3O_{10}$.

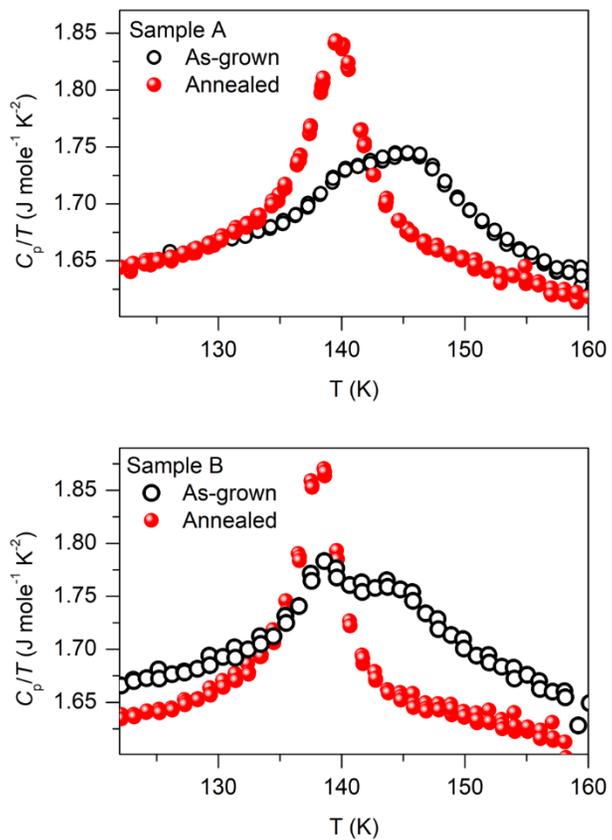

**Fig. 5** Heat capacity of two specimens of as-grown biphasic La$_4$Ni$_3$O$_{10}$ before and after annealing. See text for details.

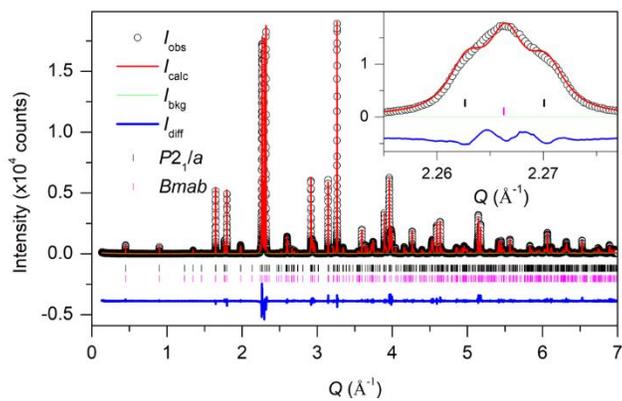

**Fig. 6** High-resolution synchrotron x-ray powder diffraction pattern and Rietveld refinement of biphasic La$_4$Ni$_3$O$_{10}$. Inserts show the quality of fit in the $Q$ range 2.255-2.277 Å$^{-1}$.

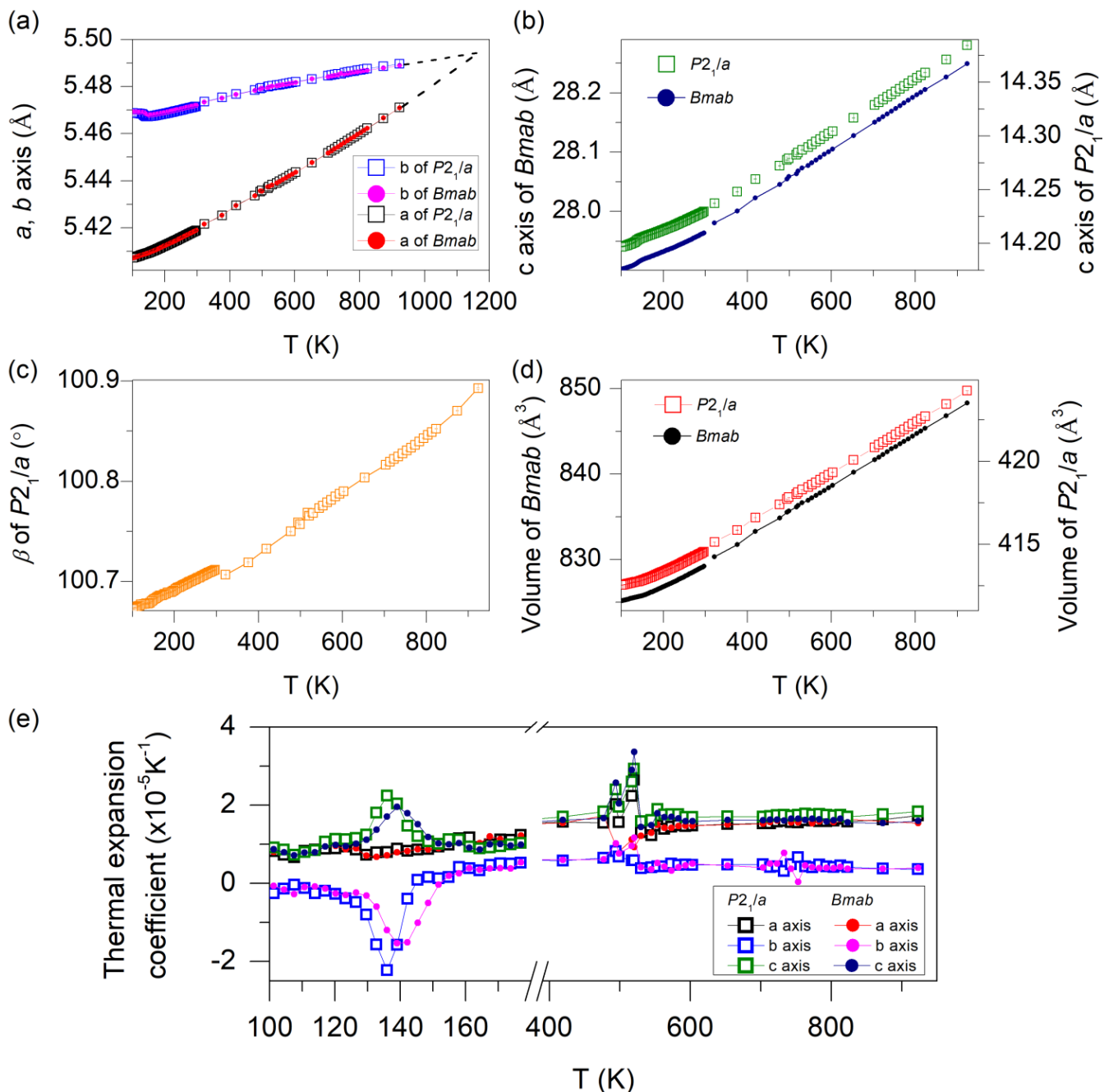

**Fig. 7** Temperature dependence of lattice parameters of La$_4$Ni$_3$O$_{10}$ from 100 K to 923 K obtained from Rietveld refinement of high-resolution synchrotron powder x-ray diffraction patterns. (a) $a$ and $b$ axes of $P2_1/a$ and $Bmab$-La$_4$Ni$_3$O$_{10}$. (b) $c$ axis. (c) $\beta$ of $P2_1/a$. (d) Volume. (e) Thermal expansion coefficients obtained from refined lattice parameters. Note the lattice parameters were obtained from two samples, one measured from 100 K to 300 K, and the other measured from 321 K to 923 K.

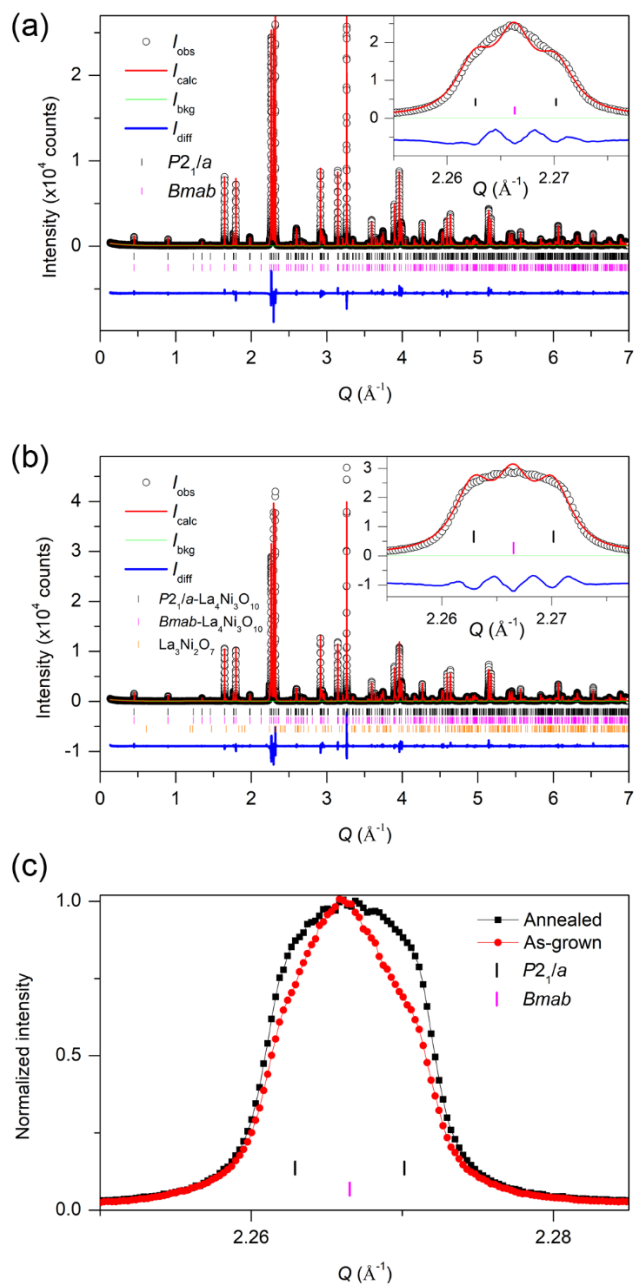

**Fig. 8** Annealing effect on biphasic $La_4Ni_3O_{10}$. (a) High-resolution synchrotron x-ray powder diffraction pattern of as-grown $La_4Ni_3O_{10}$. (b) Annealed $La_4Ni_3O_{10}$ at 1 bar $O_2$. Inserts show the quality of fit in $Q$ range 2.255-2.277 Å$^{-1}$. (c) Comparison of the peaks before and after annealing in the range of $2.25 \leq Q \leq 2.285$ Å$^{-1}$ by normalizing their heights.

**Table I**. Crystallographic data for $R_4Ni_3O_{10}$ (R=La, Pr) above the metal-to-metal transition temperatures

| Empirical formula | $La_4Ni_3O_{10}$ | $La_4Ni_3O_{10}$ | $Pr_4Ni_3O_{10}$ |
|---|---|---|---|
| Crystal system | orthorhombic | monoclinic | monoclinic |
| Space group | *Bmab* | $P2_1/a$ | $P2_1/a$ |
| Temperature, K | 200(2) | 200(2) | 298(2) |
| Formula weight | 891.77 | 891.77 | 899.77 |
| $a$ (Å) | 5.4183(5) | 5.4151(3) | 5.3816(2) |
| $b$ (Å) | 5.4696(5) | 5.4714(3) | 5.4711(3) |
| $c$ (Å) | 27.946(2) | 14.2277(7) | 14.0284(6) |
| $\beta$ (°) | 90 | 100.818(1) | 100.646(2) |
| $V$ (Å$^3$) | 828.21(12) | 414.05(4) | 405.93(3) |
| Density (calculated), g/cm$^3$ | 7.152 | 7.153 | 7.361 |
| $Z$ | 4 | 2 | 2 |
| Radiation type | Synchrotron x-ray, 0.41328 Å | Synchrotron x-ray, 0.41324 Å | Mo K$\alpha$, 0.71073 Å |
| Crystal size (mm$^3$) | 0.037 x 0.012 x 0.008 | 0.05 x 0.02 x 0.007 | 0.04 x 0.03 x 0.02 |
| Absorption coefficient, mm$^{-1}$ | 5.633 | 5.634 | 30.367 |
| Data collection Diffractometer | 15-ID-B, APS | 15-ID-D, APS | Bruker APEX2 |
| Absorption correction | Multi-Scan | Multi-Scan | Multi-Scan |
| Reflections collected | 7381 | 7359 | 4934 |
| Independent reflections | 1064 ($R_{int} = 0.0552$) | 1841 ($R_{int} = 0.0334$) | 1239 ($R_{int} = 0.0260$) |
| $\theta$ range for data collection, ° | 2.325 to 21.470 | 0.847 to 21.298 | 1.477 to 31.293 |
| F(000) | 1568 | 784 | 800 |
| Index ranges | $-8 \le h \le 8$, $-8 \le k \le 8$, $-36 \le l \le 38$ | $-9 \le h \le 9$, $-9 \le k \le 9$, $-22 \le l \le 22$ | $-7 \le h \le 7$, $-7 \le k \le 7$, $-19 \le l \le 19$ |
| Refinement method | Full-matrix least-squares on F$^2$ | Full-matrix least-squares on F$^2$ | Full-matrix least-squares on F$^2$ |
| Data / restraints / parameters | 1064 / 0 / 29 | 1841 / 0 / 79 | 1239 / 0 / 79 |
| Goodness-of-fit | 1.097 | 1.118 | 1.121 |
| $R_1/wR_2$ ($I > 2\sigma$) | 0.0595/0.1693 | 0.0284/0.0843 | 0.0341/0.0766 |
| $R_1/wR_2$ (all) | 0.0618/0.1720 | 0.0295/0.0859 | 0.0431/0.0802 |
| Largest diff. peak and hole (e Å$^{-3}$) | 6.126 and -10.383 | 2.733 and -2.209 | 3.867 and -2.353 |

**Table II**. Bond lengths of Ni-O in *Bmab*-, *P*2$_1$/*a*-La$_4$Ni$_3$O$_{10}$ and Pr$_4$Ni$_3$O$_{10}$.

| | *Bmab*-La$_4$Ni$_3$O$_{10}$ | *P*2$_1$/*a*-La$_4$Ni$_3$O$_{10}$ | Pr$_4$Ni$_3$O$_{10}$ |
|---|---|---|---|
| Inner Ni of the trilayer (Å) | 1.920(5) | 1.925(3) | 1.922(6) |
| | 1.920(5) | 1.925(3) | 1.922(6) |
| | 1.9354(9) | 1.936(4) | 1.957(7) |
| | 1.9354(9) | 1.936(4) | 1.957(7) |
| | 1.9354(9) | 1.940(4) | 1.957(7) |
| | 1.9354(9) | 1.940(4) | 1.957(7) |
| Bond valence sum | 3.01 | 2.98 | 2.89 |
| Outer Ni of the trilayer (Å) | 1.9238(6) | 1.922(3) | 1.922(6) |
| | 1.9238(6) | 1.927(3) | 1.926(6) |
| | 1.9359(6) | 1.930(3) | 1.934(6) |
| | 1.9359(6) | 1.941(3) | 1.947(6) |
| | 2.001(5) | 1.991(3) | 2.001(6) |
| | 2.165(6) | 2.170(3) | 2.154(6) |
| Bond valence sum | 2.70 | 2.70 | 2.69 |
| Ni-O-Ni angle in the ab plane (°) | 171.6(3) | 170.6(2) | 167.8(4) |
| | 171.8(3) | 172.46(19) | 165.1(4) |
| | 168.0(5) | 166.5(3) | 157.3(4) |
| Ni-O-Ni angle out of the ab plane (°) | 165.6(4) | 165.1(2) | 158.1(4) |

**Table III**. Rietveld refinement result for as-grown and annealed bi-phasic La$_4$Ni$_3$O$_{10}$.

| | As-grown bi-phasic sample | After annealing |
|---|---|---|
| | $R_{wp}$=11.0% and GOF=2.0 | $R_{wp}$=10.6% and GOF=2.3 |
| *P*2$_1$/*a*-La$_4$Ni$_3$O$_{10}$ | 59.1% mass fraction | 68.6% mass fraction |
| | *a*=5.4192 Å | *a*=5.4186 Å |
| | *b*=5.4716 Å | *b*=5.4711 Å |
| | *c*=14.2301 Å | *c*=14.2306 Å |
| | β=100.71° | β=100.72° |
| *Bmab*-La$_4$Ni$_3$O$_{10}$ | 40.9% mass fraction | 28.0% mass fraction |
| | *a*=5.4192 Å | *a*=5.4185 Å |
| | *b*=5.4721 Å | *b*=5.4712 Å |
| | *c*=27.9653 Å | *c*=27.9646 Å |
| La$_3$Ni$_2$O$_7$ | 0 | 3.4% mass fraction |
| | | *a*=5.3934 Å |
| | | *b*=5.4497 Å |
| | | *c*=20.5317 Å |

# Supplementary Materials for:

# High oxygen pressure floating zone growth and crystal structure of the metallic nickelates $R_4Ni_3O_{10}$ (R=La, Pr)

Junjie Zhang,[1,2,3]* Hong Zheng,[1] Yu-Sheng Chen,[4] Yang Ren,[5] Masao Yonemura,[6,7] Ashfia Huq,[8] and J. F. Mitchell[1]*

[1]Materials Science Division, Argonne National Laboratory, Argonne, IL 60439, United States

[2]Materials Science and Technology Division, Oak Ridge National Laboratory, Oak Ridge, TN 37831, United States

[3]Institute of Crystal Materials, Shandong University, Jinan, Shandong 250100, China

[4]ChemMatCARS, The University of Chicago, Argonne, IL 60439, United States

[5]X-ray Science Division, Advanced Photon Source, Argonne National Laboratory, Argonne, IL 60439, United States

[6]Institute of Materials Structure Science, High Energy Accelerator Research Organization, 1-1 Oho, Tsukuba, Ibaraki 305-0801, Japan

[7]Sokendai (The Graduate University for Advanced Studies), Shirakata 203-1, Tokai, Naka 319-1106, Japan

[8]Neutron Scattering Science Directorate, Oak Ridge National Laboratory, Oak Ridge, Tennessee 37831, United States

*email: junjie@sdu.edu.cn; mitchell@anl.gov

.

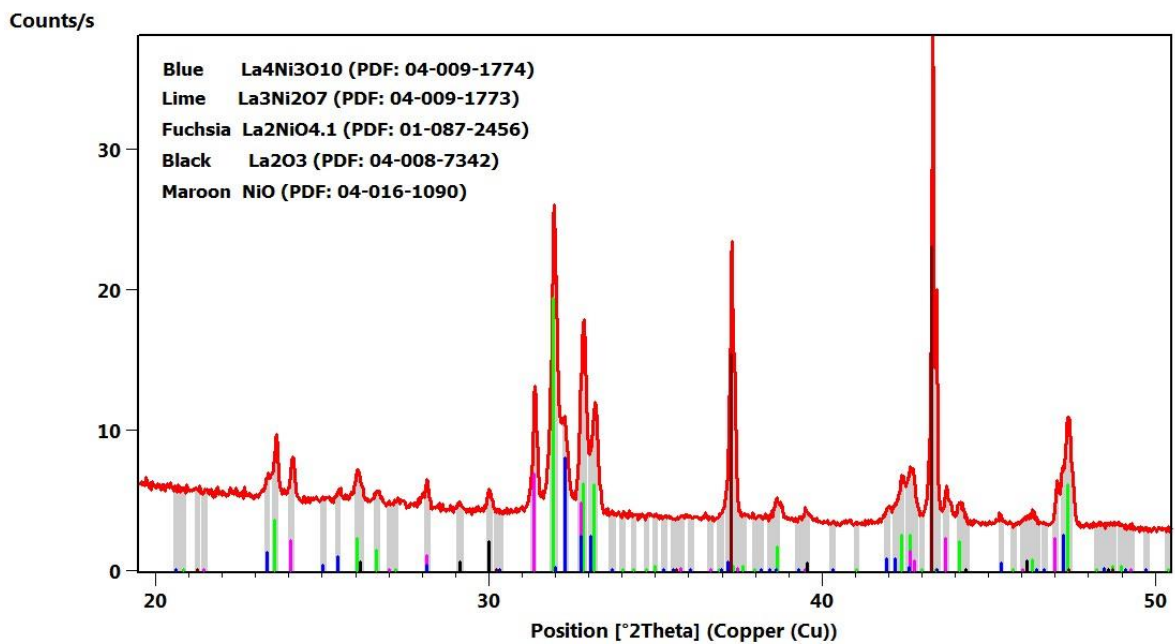

**Fig. S1** Lab x-ray diffraction pattern of precursor of nominal composition La$_4$Ni$_3$O$_{10}$.

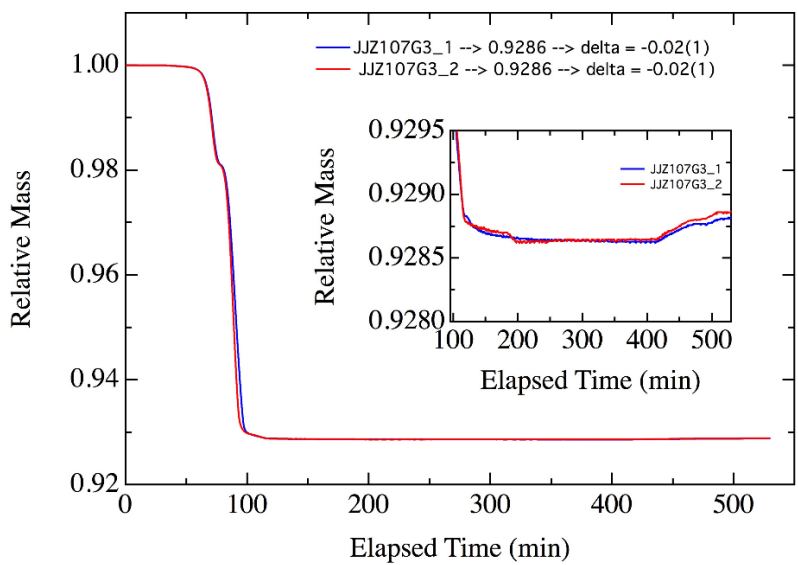

**Fig. S2** H$_2$-N$_2$ reduction process of La$_4$Ni$_3$O$_{10}$; similar data are obtained for Pr$_4$Ni$_3$O$_{10}$ (not shown). A clear endpoint is seen after ~250-300 min at 900 °C in this atmosphere. The small mass uptake after 400 min occurs during cooling and may reflect either an error in buoyancy correction or an uptake of moisture from the environment. We calculate the O content from the plateau/endpoint mass. Two experiments here using sample from the same region of the boule are shown to demonstrate reproducibility of the technique. It is possible that O content varies across the boule, but we did not examine this.

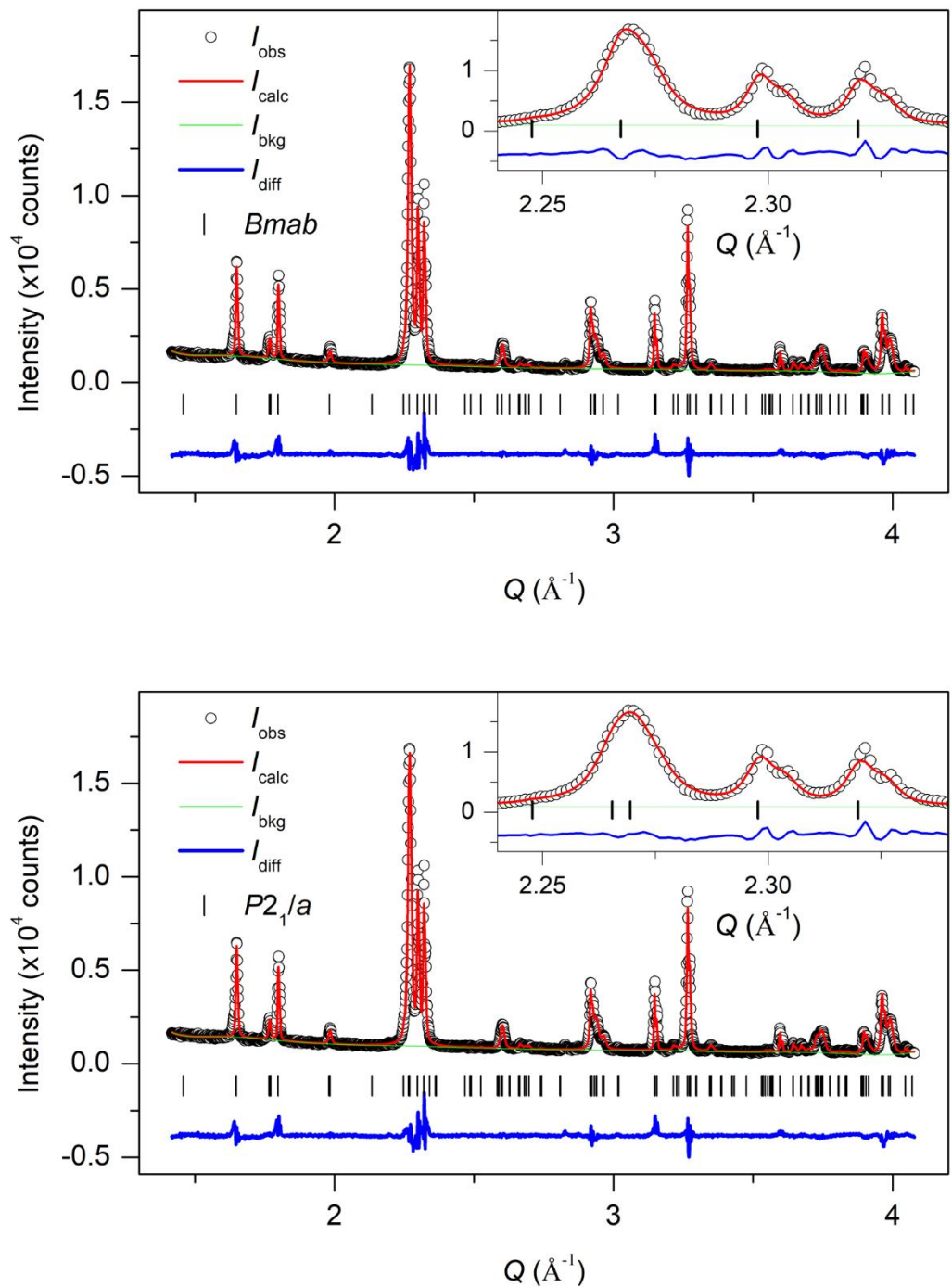

**Fig. S3** Lab x-ray diffraction pattern of as-grown $La_4Ni_3O_{10}$ and Rietveld refinement using *Bmab* and *P2$_1$/a*. Both refinements converged to $R_{wp}$=7.5% and GOF=2.7. Inset shows the quality of fit for 2.24 Å$^{-1}$≤$Q$ ≤2.34 Å$^{-1}$.

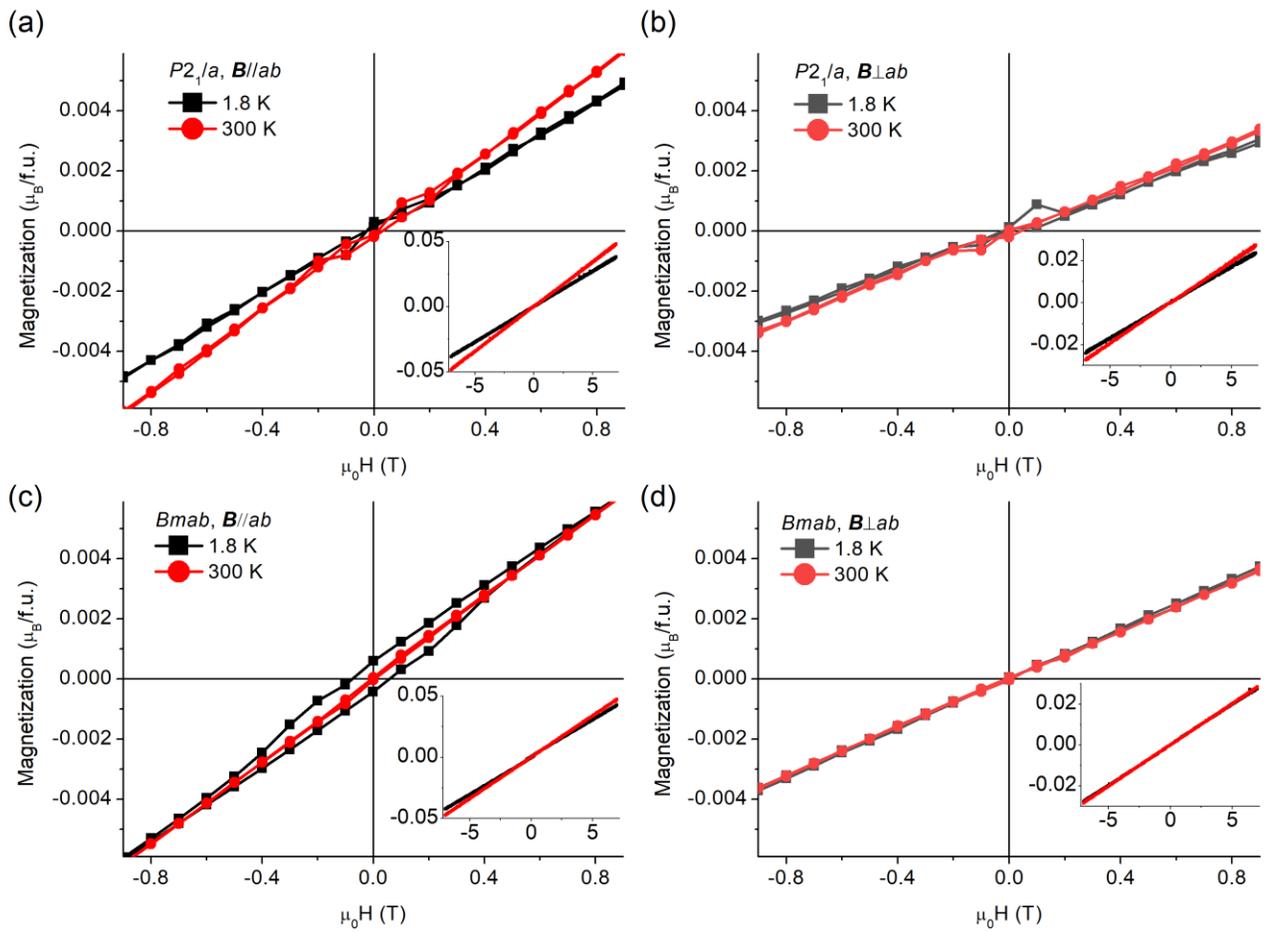

**Fig. S4** Isothermal magnetization as a function of magnetic field for La$_4$Ni$_3$O$_{10}$.

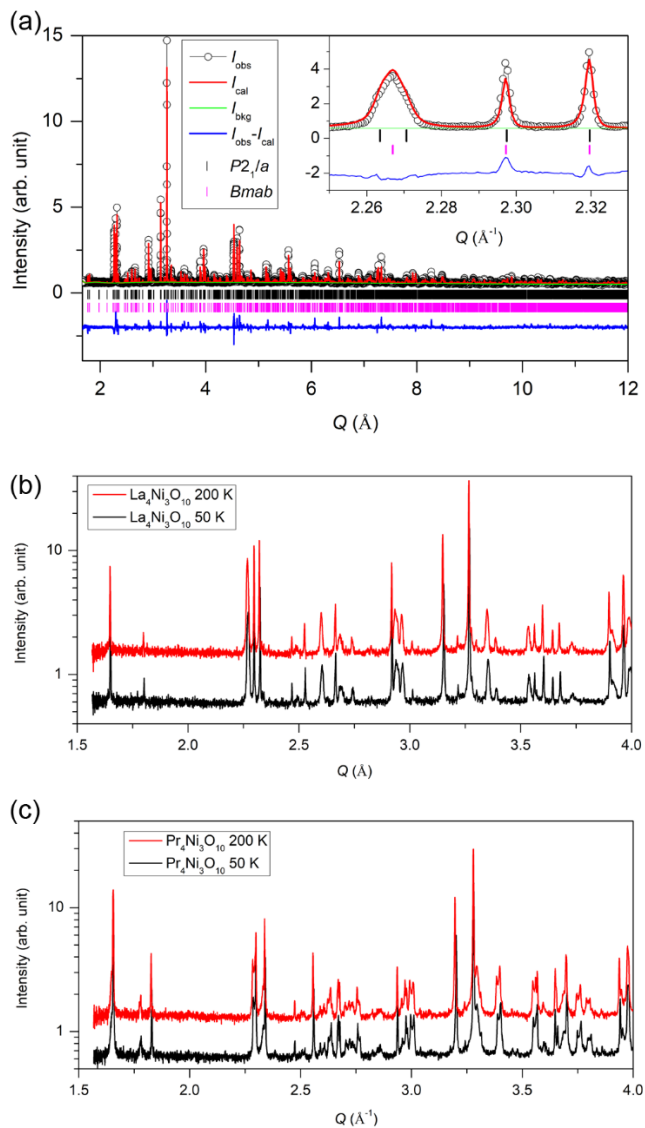

**Fig. S5** (a) High-resolution time-of-flight neutron powder diffraction pattern and Rietveld analysis of biphasic $La_4Ni_3O_{10}$ at room temperature. (b,c) High-resolution time-of-flight neutron powder diffraction patterns of bi-phasic $La_4Ni_3O_{10}$ and $Pr_4Ni_3O_{10}$ at 200 and 50 K.

**High-Resolution Neutron Powder Diffraction.** To confirm the structure determined from x-rays and also to investigate any potential magnetic order at low temperature, high-resolution neutron powder diffraction experiments were carried out on $Pr_4Ni_3O_{10}$ and biphasic $La_4Ni_3O_{10}$ powder samples on a time-of-flight (TOF) Super High Resolution Powder Diffractometer, Super HRPD [1,2] at the Material and Life science Facility (MLF) in the Japan Proton Accelerator Research Complex (J-PARC). The high-resolution neutron powder diffraction of biphasic $La_4Ni_3O_{10}$ at room temperature with Rietveld refinement using a mixed-phase *Bmab* and *P2₁/a* model is shown in Fig. S5(a). Diffraction data were also collected at 200 and 50 K, and Fig. S5(b,c) show the respective diffraction patterns. As can be seen, no extra peaks attributable to magnetic order were found at 50 K, demonstrating that either no long-range antiferromagnetic order occurs across the metal-to-metal transitions or that the ordered moment is below the detection limit of the experiment. We have recently performed single crystal diffraction on $R_4Ni_3O_{10}$ (R=La, Pr) crystals and found spin-density-wave order in these metallic trilayer nickelates [3].

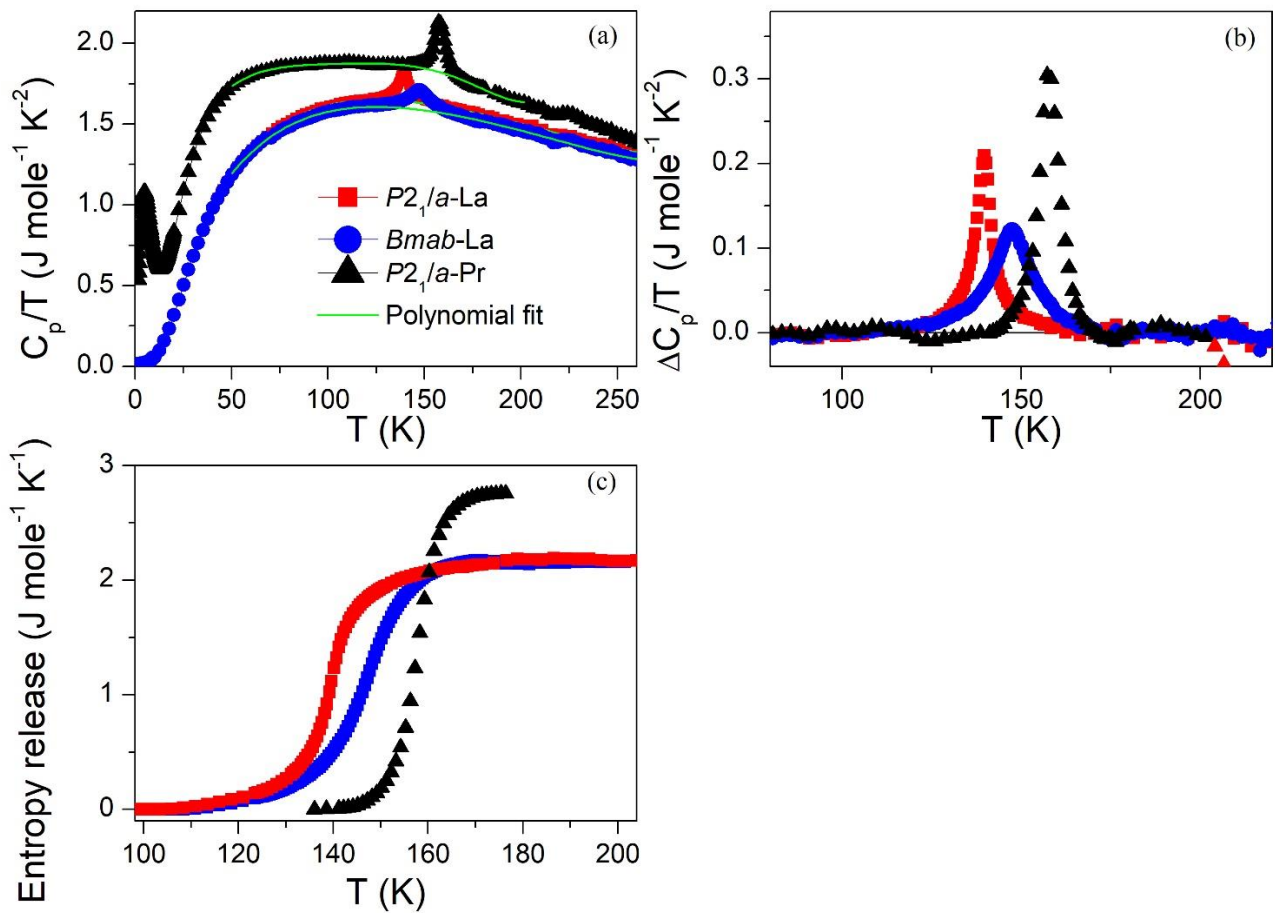

**Fig. S6** Heat capacity of $R_4Ni_3O_{10}$ (R=La and Pr). (a) $C_p/T$ as a function of temperature. Green curves denote background estimated by polynomial fit. (b) Difference of $C_p/T$ and background. (c) Entropy change across the phase transition.

**Table S1.** Unit cell setting, R factor using isotropic thermal parameters and systematic absence violations/inconsistent equivalents for monoclinic $R_4Ni_3O_{10}$ (R=La, Pr).

| Compound | Sample ID | Unit cell setting a, b, c, β | R factor (Isotropic thermal parameters) | Systematic absence violations/Inconsistent equivalents |
|---|---|---|---|---|
| $La_4Ni_3O_{10}$ | JJZ101G3_50by20by7um_200K | 27.9502, 5.4713, 5.4150, 90.139 | $R_1$=8.26%, $wR_2$=26.29% | 13 violations (h0l, h=-6, -10, -10, -12, -12,-16, -26, 26, l=1) 0 Inconsistent equivalents |
| | | 14.2277, 5.4714, 5.4151, 100.818 | $R_1$=5.78%, $wR_2$=18.65% | 0/0 |
| | JJZ107G3_30by30by10um_200K | 27.9456, 5.4700, 5.4136, 90.06 | $R_1$=6.87%, $wR_2$=31.66% | 4 violations (h0l, h=-10, -12, -16, 26 and l=1) 0 Inconsistent equivalents |
| | | 14.2308, 5.4702, 5.4141, 100.911 | $R_1$=6.17%, $wR_2$=21.05% | 0/0 |
| $Pr_4Ni_3O_{10}$ | JJZ103G2_24by14by10um_200K | 27.5089, 5.4621, 5.3790, 90.405 | $R_1$=11.3%, $wR_2$=40.4% | 30 violations (h0l, l should be even, but l=1,3,5, h=-4,-6,-8...) 4 Inconsistent equivalents |
| | | 14.0152, 5.4675, 5.3817, 100.748 | $R_1$=13.8%, $wR_2$=42.41% | 0/0 |

**Synchrotron x-ray single crystal diffraction.** Data were collected on single crystals of monoclinic $La_4Ni_3O_{10}$ at 200 K at Beamline 15-ID-D at the Advanced Photon Source. Data were reduced using monoclinic settings with $c$~27.95 Å and $c$~14.2 Å to find out the best solution. Table S1 lists the sample ID, unit cell setting, $R$ factor using isotropic thermal parameters and systematic absence violations/inconsistent equivalents for both solutions. As can be seen clearly, the solution using unit cell setting with $c$~27.95 Å has multiple systematic absence violations, thus it is not appropriate to describe the symmetry of the monoclinic phase. Indeed, the solution using $c$~14.2 Å shows no violations/inconsistent equivalents and better $R$ factors. The same situation was found for $Pr_4Ni_3O_{10}$. Therefore, we choose the unit cell setting with $c$~14.2 Å to describe the structure of the monoclinic $R_4Ni_3O_{10}$ (R=La, Pr).

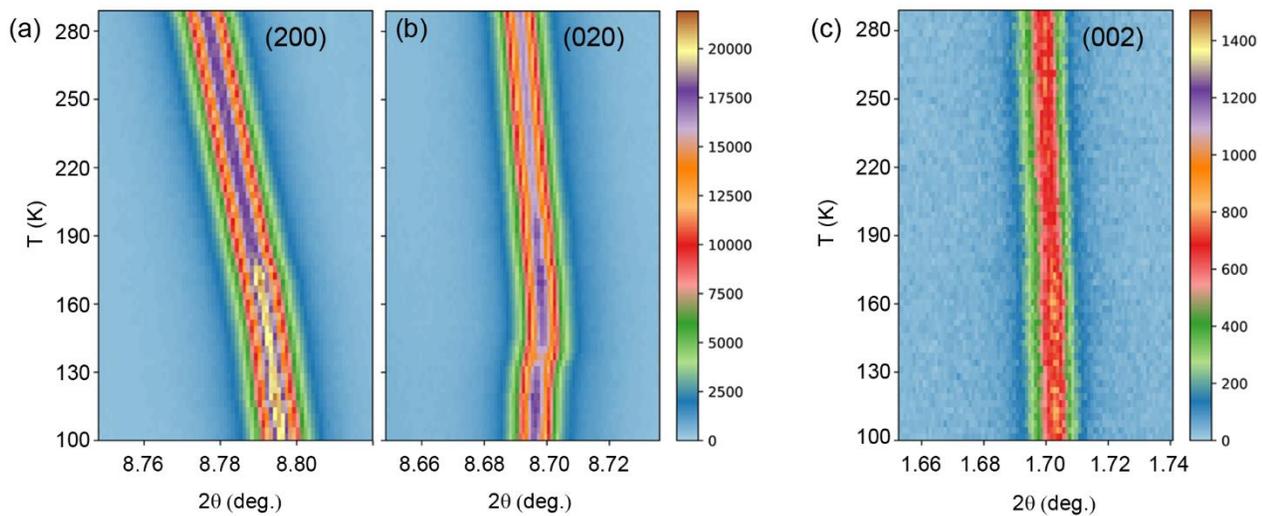

**Fig. S7** Evolution of the (200), (020) and (002) peaks as a function of temperature of biphasic $La_4Ni_3O_{10}$.

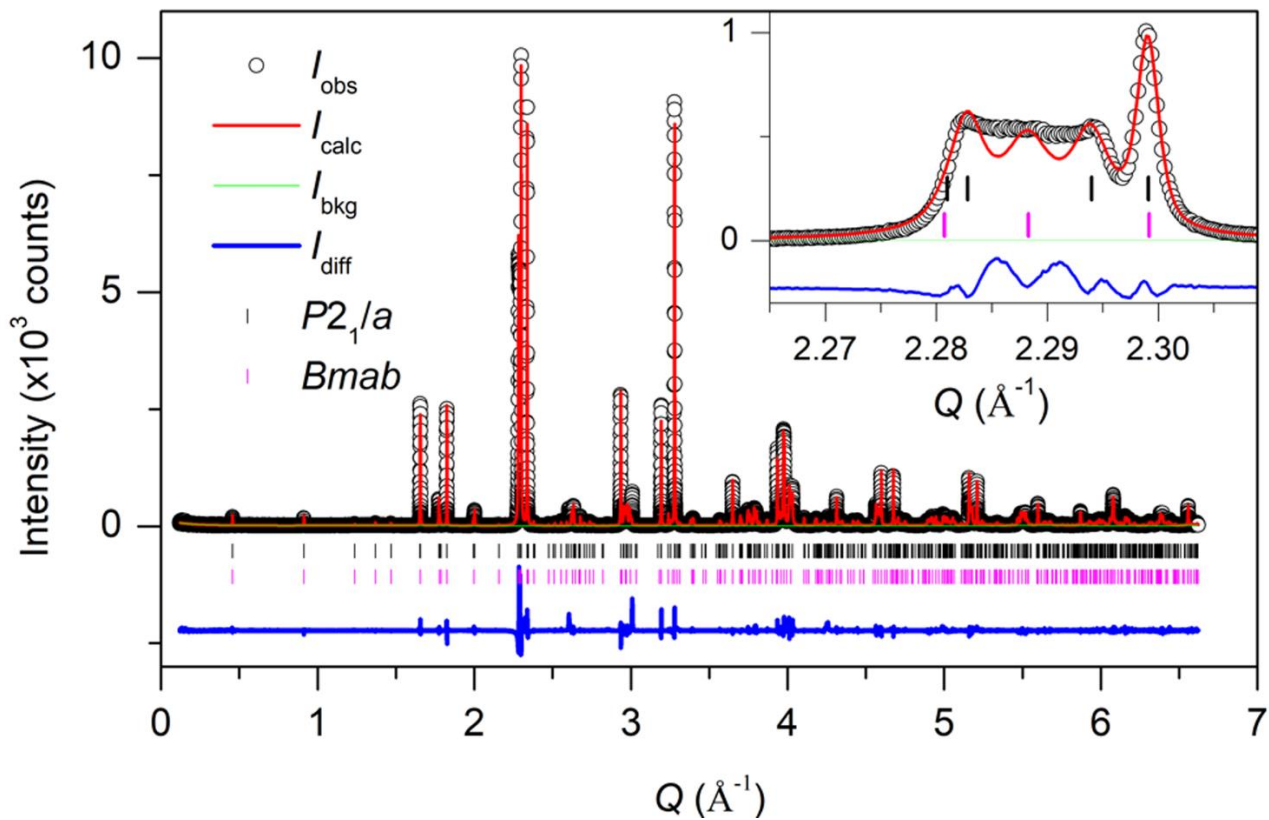

**Fig. S8** High-resolution synchrotron powder x-ray diffraction pattern of $Pr_4Ni_3O_{10}$ used for in-situ experiments.

**High-resolution synchrotron x-ray powder diffraction of $Pr_4Ni_3O_{10}$.** Significant peak broadening from the anisotropic strain and stacking faults has been observed in the patterns of $Pr_4Ni_3O_{10}$, as reported in Ref. [4,5]. To simulate the intensity between the peaks of monoclinic phase, we modeled the patterns using a combination of $P2_1/a$ and a putative $Bmab$ phase. For the $Bmab$ phase, we adopt the CIF file of $Bmab$-$La_4Ni_3O_{10}$ as a starting point. The Rietveld refinement, shown in Fig. S8, converged to $R_{wp}$=18.8%, GOF=1.9 with a mass fraction of 36.6%wt for $Bmab$ and 63.4%wt for $P2_1/a$. The obtained lattice parameters are $a$=5.3760 Å, $b$=5.4655 Å, $c$=27.5487 Å for $Bmab$, and $a$= 5.3761 Å, $b$= 5.4657 Å, $c$= 14.0148 Å and $\beta$=100.67° for $P2_1/a$.

Fig. S9 presents the temperature dependence of the unit cell parameters $a$, $b$, $c$, $\beta$, and $V$ extracted from Rietveld refinements, as well as the calculated linear thermal expansion coefficient along $b$ axis. An anomaly at ~157.6 K is observed for both phases. As shown in the main text for the case of $La_4Ni_3O_{10}$, $P2_1/a$ and $Bmab$ phases have distinct transition temperatures. Thus, the appearance of the lattice anomaly at the same temperature for both $P2_1/a$ and putative $Bmab$ phases is strong evidence that they are actually the same phase and the intensity that appears between the monoclinic peaks comes not from a $Bmab$ phase but rather from defects such as stacking faults. The modeling of such defects is beyond the scope of this paper.

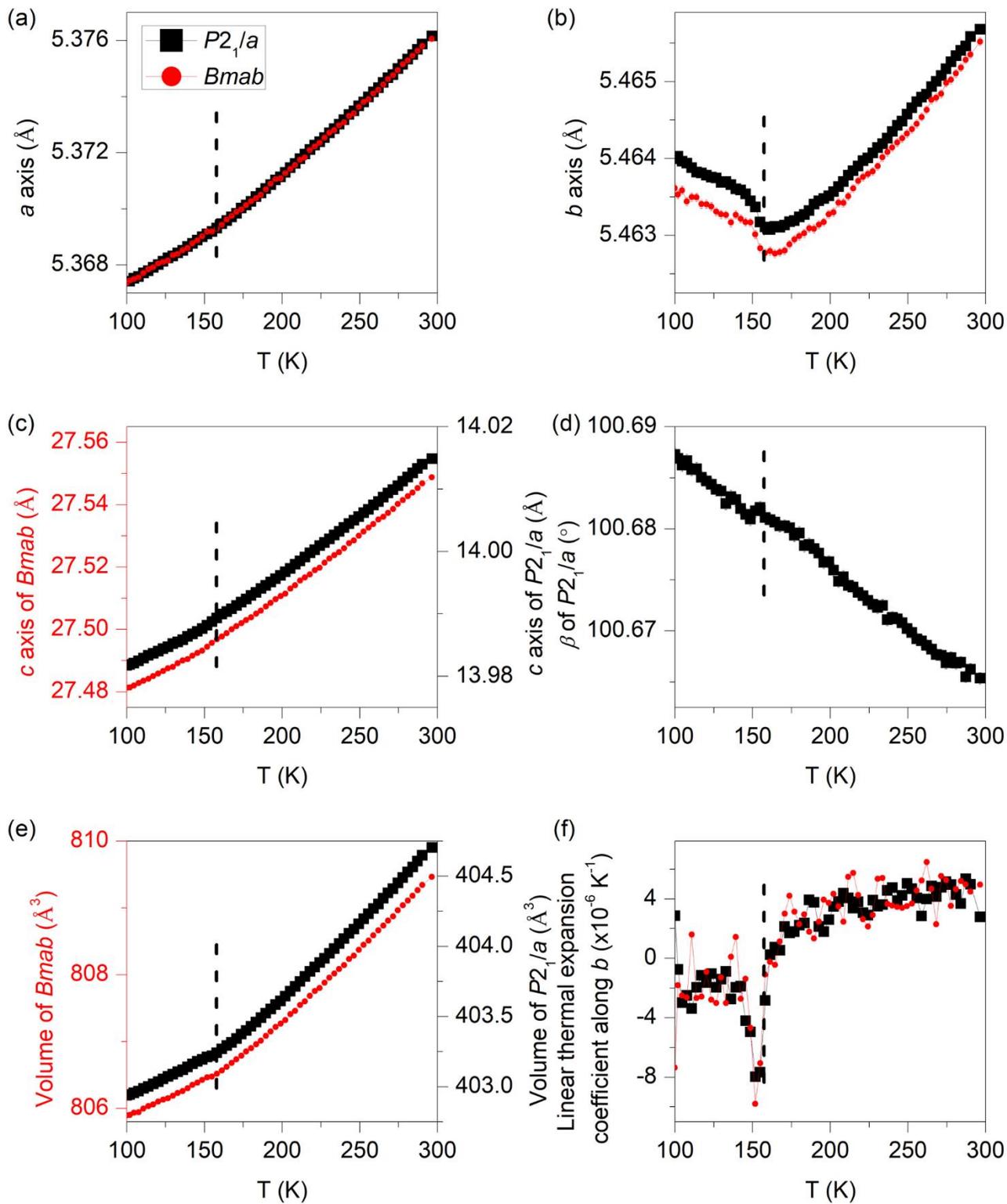

**Fig. S9** Temperature dependence of lattice parameters of $Pr_4Ni_3O_{10}$ on cooling.

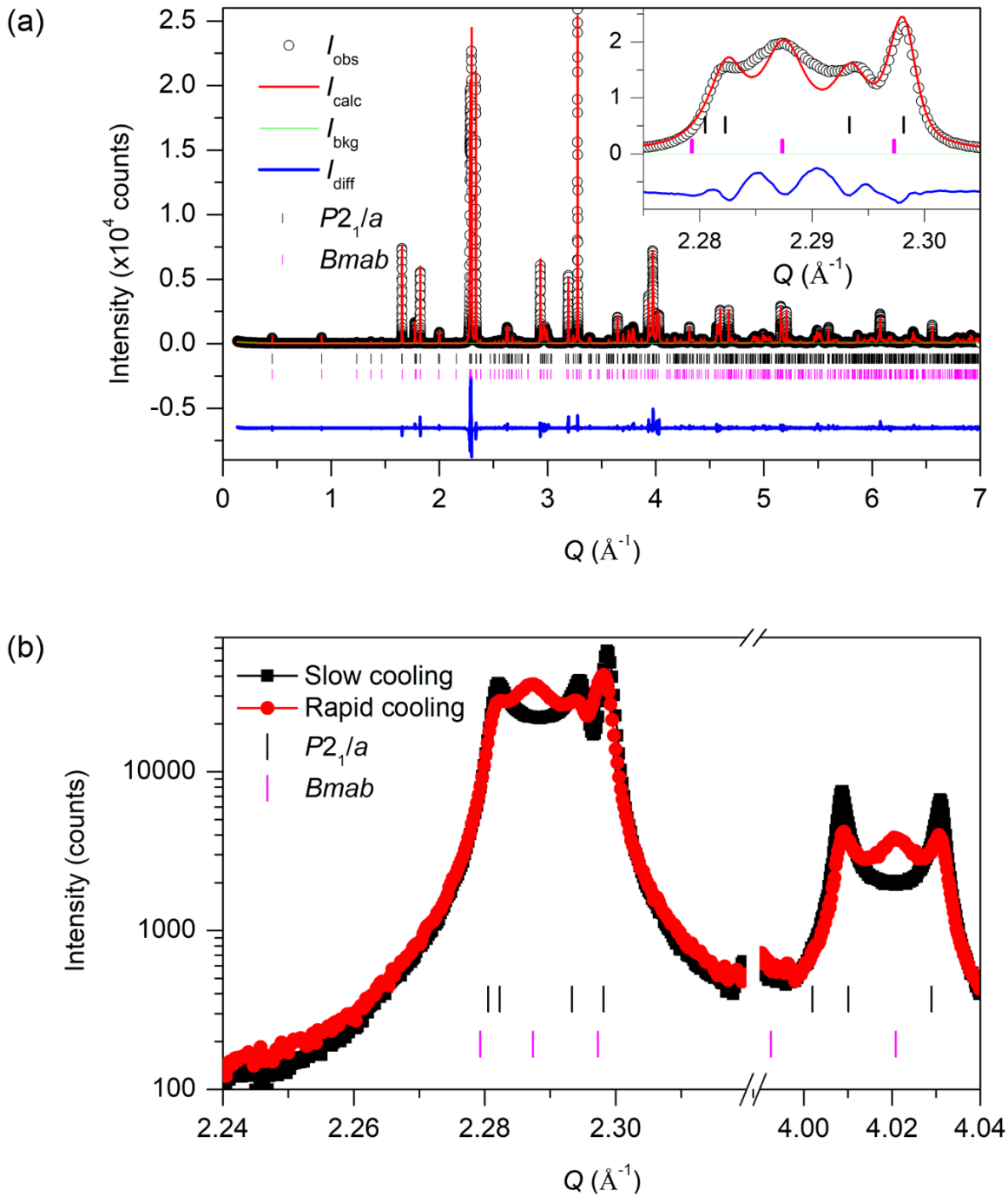

**Fig. S10** High-resolution synchrotron powder x-ray diffraction pattern of $Pr_4Ni_3O_{10}$ using different post-growth cooling procedures. (a) Rietveld refinement of the sample taken from the end of as-grown boule after rapid cooling ($\lambda$=0.413367 Å, 11-BM, APS). $R_{wp}$=13.7%, GOF=2.36. Inset shows the quality of fit in the range 2.24 Å$^{-1}$≤$Q$≤2.34 Å$^{-1}$. Obtained lattice parameters are: $a$=5.3741 Å, $b$=5.4700 Å, $c$=27.5660 Å for $Bmab$-$Pr_4Ni_3O_{10}$, and $a$=5.3773 Å, $b$=5.4680 Å, $c$=14.0181 Å and $\beta$=100.67° for $P2_1/a$-$Pr_4Ni_3O_{10}$. (b) Comparison between samples of slow cooling and quenching after growth.

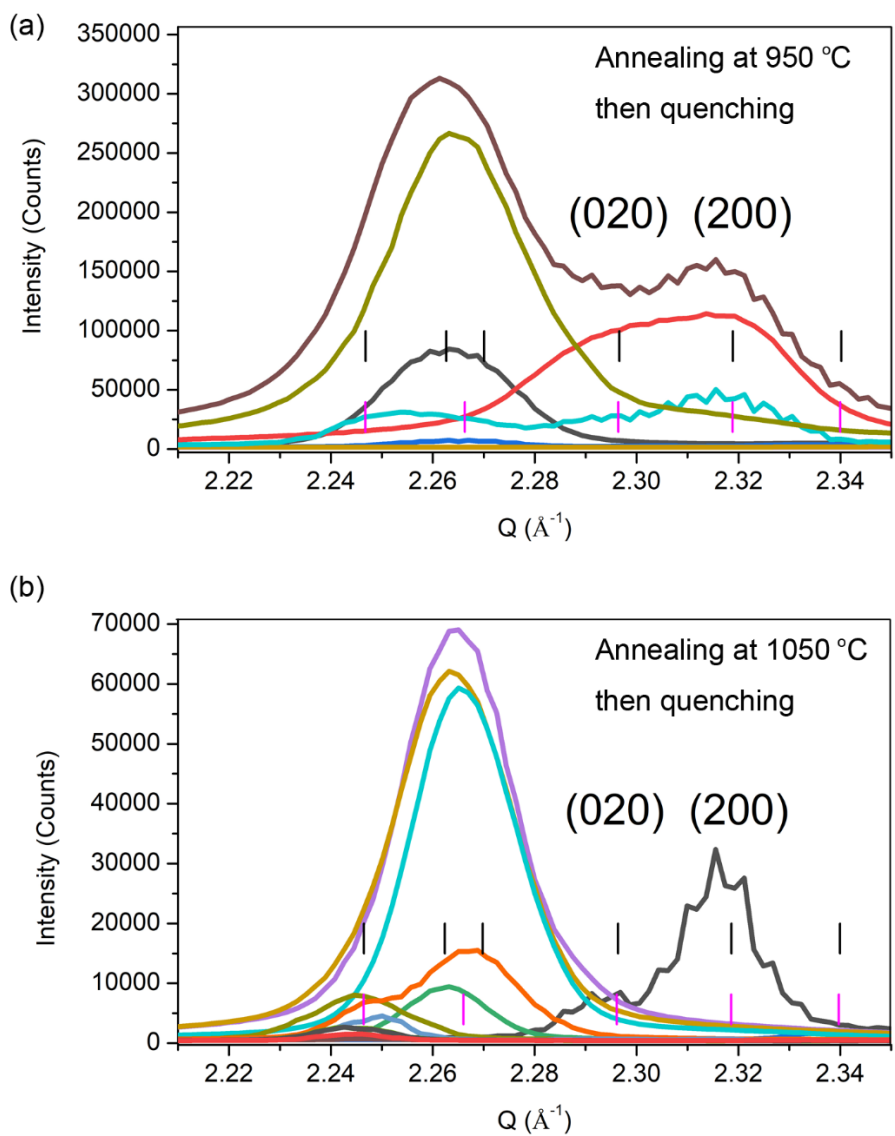

**Fig. S11** Synchrotron x-ray single crystal diffraction patterns collected by rotating crystals for 10 degrees at 0, 10, 20, 30, 40, 50, 60, 70, 80, 90°. The superposition of all patterns is comparable to the powder diffraction pattern. (a) Diffraction patterns obtained from sample quenching from 950 °C at 20 bar $O_2$; (b) Diffraction patterns from quenching at 1050 °C at 20 bar $O_2$. The ticks indicate positions of peaks of $P2_1/a$ (upper) and $Bmab$ (lower).

**Additional annealing experiments.** In an attempt to isolate the tetragonal phase reported by Nagell *et al*. [4] and Amow *et al*. [6], we performed additional annealing experiments on biphasic $La_4Ni_3O_{10}$ single crystals under $pO_2$=20 bar, where we quenched the crystals from 1050 or 950 ℃. Despite several attempts, we did not find a signature of the tetragonal phase – the merging of (020) and (200) peaks as reported by Nagell *et al*. [4] Fig. S11(a) and Fig. S11(b) show the diffraction patterns of the annealed bulk crystals at various angles by rocking 10 degrees at the beamline 11-ID-C at the Advanced Photon Source. Continuous rotating results in a diffraction pattern comparable to powder. As can be seen, the annealed crystals still show separated (020) and (200) peaks at ~2.295 and ~2.32 Å⁻¹, respectively, indicating the lack of a tetragonal phase.